\documentclass[12pt,preprint]{aastex}

\shorttitle{Helium abundance during a C-class flare}
\shortauthors{Andretta et al.}

\newcommand\NHe{\ensuremath{N_\mathrm{He}}}
\newcommand\NH{\ensuremath{N_\mathrm{H}}}

\newcommand\Ne{\ensuremath{N_\mathrm{e}}}
\newcommand\Pe{\ensuremath{P_\mathrm{e}}}

\newcommand\Teff{\ensuremath{T_\mathrm{eff}}}

\newcommand\logAbHe{\mbox{[He/H]}}
\newcommand\AbHe{\ensuremath{A_\mathrm{He}}}

\newcommand\Lalpha{\mbox{Ly-}\ensuremath{\alpha}}
\newcommand\Halpha{\mbox{H}\ensuremath{\alpha}}
\newcommand\CaK{\mbox{\ion{Ca}{2}~K}}

\newcommand\Iref{\ensuremath{I_\mathrm{ref}}}
\newcommand\Icor{\ensuremath{I_\mathrm{cor}}}
\newcommand\DEM{\mbox{DEM}}

\newcommand{\SEM}[1][]{\ensuremath{\mathrm{SEM}_{#1}}}

\begin{document}



\title%
{Helium line formation and abundance during a C-class flare}


\author{Vincenzo Andretta}
\affil{INAF - Osservatorio Astronomico di Capodimonte, Salita Moiariello 16,
  I-80131 Napoli, Italy}
\email{andretta@oacn.inaf.it}

\author{Pablo J. D. Mauas}
\affil{Instituto de Astronom\'\i a y F\'\i sica del Espacio, Argentina}

\author{Ambretta Falchi}
\affil{INAF - Osservatorio Astrofisico di Arcetri, Italy}

\author{Luca Teriaca}
\affil{Max-Planck-Institut f\"ur Sonnensystemforschung, Germany}



\begin{abstract}
  {During a coordinated campaign which took place in May 2001, a
    C-class flare was observed both with SOHO instruments and with the Dunn
    Solar Telescope of the National Solar Observatory at Sacramento Peak. In
    two previous papers we have described the observations and discussed some
    dynamical aspects of the earlier phases of the flare, as well as the
    helium line formation in the active region prior to the event.  Here we
    extend the analysis of the helium line formation to the later phases of
    the flare in two different locations of the flaring area.  We have devised
    a new technique, exploiting all available information from various SOHO
    instruments, to determine the spectral distribution of the photoionizing
    EUV radiation produced by the corona overlying the two target regions.  In
    order to find semiempirical models matching all of our observables, we
    analyzed the effect on the calculated helium spectrum both of \AbHe\ (the
    He abundance) and of the uncertainties in the incident EUV radiation
    (level and spectral distribution).  We found that the abundance has in
    most cases (but not in all) a larger effect than the coronal
    back-radiation.  The result of our analysis is that, considering the error
    of the measured lines, and adopting our best estimate for the coronal EUV
    illumination, the value $\AbHe=0.075\pm 0.010$ in the chromosphere (for
    $T>6300$~K) and transition region yields reasonably good matches for all
    the observed lines.  This value is marginally consistent with the most
    commonly accepted photospheric value: $\AbHe=0.085$.}
\end{abstract}






\section{Introduction}\label{sec:intro}

A key ingredient to understand the physics of solar and stellar plasmas is
their chemical composition. An astrophysical plasma is subject to chemical
fractionation processes of diverse nature \citep[gravity, thermal diffusion,
etc., see e.~g.][] {Drake:03} that can produce abundance ``anomalies'' between
regions of different temperature and density. For the solar corona these
anomalies are generally described in terms of the so-called FIP (\textit{First
  Ionization Potential}) effect.  Not only the details, but the very general
physical framework behind this effect is still debated.  However, one of the
few aspects over which there is apparently a consensus, is that the FIP effect
should arise in the chromosphere \citep{Geiss:82,Geiss:98}.

Helium is one of the few elements that exhibits strong lines forming in the
chromosphere, and thus, in principle at least, its abundance ($\logAbHe \equiv
\log\AbHe$ (where $\AbHe \equiv \NHe/\NH$, the ratio of number densities of He
and H) could be estimated in that region.  It is also the element with the
highest FIP (24~eV).  Furthermore, it is the second most abundant element in
the Sun (and in the universe, for that matter), and therefore must be included
in any theoretical model of fractionation processes in the solar atmosphere.
Moreover, the coronal \logAbHe\ strongly depends on the availability of helium
in the upper chromosphere 
\citep[e.~g.:][]{Killie-etal:05}.

Unfortunately, analyzing helium lines in the solar atmosphere is a
particularly challenging task, more so than other ``typical'' chromospheric
lines such as \Halpha\ or the \ion{Ca}{2}~H~\&~K doublet. The main reasons are
described in several papers \citep[for
example:][]{Andretta-Jones:97,Pietarila-Judge:04}. Among the problems that can
be mentioned: all of its lines need detailed radiative transfer calculations;
strong interlocking between $n=2$ and $n=3$ singlet and triplet levels makes
the \emph{simultaneous} treatment of EUV and optical lines almost unavoidable;
the high excitation energies of all helium lines (a consequence of its high
FIP value) makes those lines very sensitive to non-equilibrium effects; the
ionization rates of neutral and ionized helium can be significantly affected
by EUV radiation via the so-called ``Photoionization - Recombination'' (P-R)
mechanism \citep[e.~g.:][]{Zirin:75,Andretta-etal:03}, and therefore not only
coronal plasmas, but even the strong \ion{He}{2} \Lalpha\ at 30.4~nm can be
significant photoionization sources for neutral helium (another example of the
strong interlocking effects in the helium atomic system).

All these problems imply, among other things, that a realistic study of helium
lines in the solar chromosphere requires quite sophisticated modeling, as well
as simultaneous measurements in both the EUV and the optical (visible and IR)
range.  To this aim we planned an observing campaign (SOHO Joint Observing
Programme 139) coordinated between ground based and SOHO instruments to obtain
simultaneous spectroheliograms of the same area in several spectral lines,
including four He lines (\ion{He}{1} 58.76, \ion{He}{1} 1083.0, \ion{He}{1}
58.4 and \ion{He}{2} 30.4~nm), that sample the solar atmosphere from the
chromosphere to the transition region.  The EUV radiation in the range
$\lambda<50$~nm and in the range $26<\lambda<34$~nm has also been measured at
the same time.

In a previous paper {\citep*[][hereafter Paper~II]{Mauas-etal:05}},
we carried out an analysis of the helium line formation in a specific location
of the observed active region (NOAA AR 9468, $\cos\theta$=0.99), prior to a
small two-ribbon flare (GOES class C1) developed in this region on 2001, May
26, around 16:00~UT.  We found that the incident coronal radiation has a
limited effect on the UV He lines \citep[in partial agreement with the
proxy-based analysis of][of the \ion{He}{1} 58.4 line]{Fredvik-Maltby:99},
while being fundamental for the optical lines.  For these lines, we also found
that photons of shorter wavelengths are more effective at increasing their
line depth. Incidentally, this latter finding implies that for the calculation
of the He line profiles it is important, at least in principle, to correctly
estimate not only the total number of ionizing photons, but also their
spectral distribution.  But the main result of Paper~II was that the
photospheric value of \logAbHe\ is compatible with our observations within an
uncertainty of about a factor two and that such uncertainty could be reduced
in a higher density regime.

On the basis of these results we decided to extend our analysis to the data
taken during the flare.  In this paper we present results of such analysis and
study the formation of He lines after the main evaporation phase, when the He
lines presumably originate in a high density plasma and the incident coronal
radiation, larger than in the pre-flare phase, can be estimated with a lower
uncertainty.  We construct semiempirical models of the flaring atmosphere to
match the observed line profiles from the chromosphere to the transition
region, taking into account the total EUV irradiance in the range
$\lambda<50$~nm.  We also analyze the effect of \logAbHe\ changes, with the
purpose of attaining an empirical estimate of that parameter.

To this regard, we mention that techniques based on $\gamma$-ray line
measurements produced in flares offer an alternative avenue for estimating the
chromospheric He abundance
\citep[e.~g.][]{Mandzhavidze-etal:97,Mandzhavidze-etal:99,Share-Murphy:98}.
Results from these techniques seem generally to indicate an enhancement 
either of the accelerated $\alpha$/proton ratio or of the ``ambient'' He
abundance. Attempts at resolving this ambiguity seem however to hint that
\AbHe\ is consistent with its accepted photospheric values
\citep{Murphy:07}.

For completeness, we should also mention recent EUV spectroscopic measurements
of \AbHe\ during a flare, made by analyzing the
Balmer-$\gamma$ of \ion{He}{2} at 108.5~nm in off-limb spectra
\citep{Feldman-etal:05}, which yielded $\AbHe\sim 0.122 \pm 0.024$,
marginally higher than the photospheric value.  This result, however, refers
to off-limb post-flare loops and thus cannot be directly compared with our
on-disk, chromospheric measurements.

In Sec.~\ref{sec:obs} we briefly describe the set of observations, focussing
our attention on two specific locations of the flaring region.  In
Sec.~\ref{sec:models:jcor} we describe the procedures employed to estimate the
incident coronal EUV radiation illuminating the regions under examination and
in Sec.~\ref{sec:models:atmo} the chromospheric modeling for those two
locations.  Finally, in Sec.~\ref{sec:disc} we explore how the helium spectrum
is affected by the value of the chromospheric He abundance and by the
uncertainties in the estimates of coronal radiation.

\section{The Observations}\label{sec:obs}

A detailed description of the observing program is given in
{\cite*{Teriaca-etal:03}} (hereafter Paper~I).  We recall here that
spectroheliograms were acquired with the Horizontal Spectrograph at the Dunn
Solar Telescope (DST) of the National Solar Observatory at Sa\-cra\-men\-to
Peak in the chromospheric lines \CaK, \Halpha, and \ion{Na}{1}~D as well as in
the \ion{He}{1} lines at 587.6 (D$_3$) and 1083 nm.  The full field of view
(FOV) of the DST, $170\arcsec\times 170\arcsec$, was covered in about 5
minutes, with a sampling step of $2\arcsec$. Correcting for offsets among the
different detectors resulted in a final useful FOV of $160\arcsec\times
140\arcsec$ with an effective resolution of $2\arcsec$.

During the same period, spectroheliograms of the active region were obtained
in selected spectral lines with the Normal Incidence Spectrometer (NIS) of the
Coronal Diagnostic Spectrometer (CDS) \citep{Harrison-etal:95} aboard SOHO.
The selected lines are: \ion{Fe}{16}~36.1~nm in the NIS band \#1;
\ion{He}{1}~58.4~nm, \ion{He}{2}~30.4~nm (2$^\mathrm{nd}$ order), \ion{O}{5}
62.9~nm, and the blend \ion{Fe}{12}~59.26 + \ion{Fe}{19}~59.22~nm%
\footnote{%
  In the case of the \ion{Fe}{12}~59.26+\ion{Fe}{19}~59.22~nm blend, an
  analysis of the mean line wavelengths shows (see also Paper~I) that the
  former contribution is small or even negligible in the areas interested by
  the flare, but, conversely, is dominant in the surrounding, unperturbed
  regions.%
}
in the NIS band \#2.  The $4\arcsec$-wide slit was stepped $6\arcsec$ covering
a $148\arcsec$ wide area in $\sim 5.5$ minutes.  The final useful FOV was
$148\arcsec\times 138\arcsec$ with an effective pixel size of $6\arcsec\times
3\arcsec.4$.  Ground-based and CDS data were aligned using SOHO/MDI images as
a reference.  We estimate the error around few arcseconds.  CDS and ground
based spectra are simultaneous within 2 minutes.
 
\begin{figure*}[h!tb]
  \centering
  \includegraphics%
  [angle=90,width=1.0\linewidth]%
  {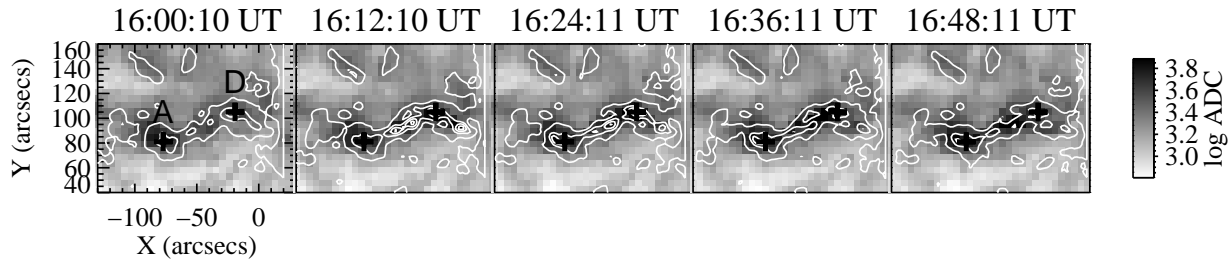}
  \caption{%
    {The} evolution of the NOAA AR
    9468 in the first hour of the
    event, as seen in the sequence of SOHO/EIT 19.5~nm images (see
    Fig.~\protect\ref{fig:hxr} for a comparison with XUV
    irradiance and \Halpha\ radiance light curves). The contours represent 
    \Halpha\ isophotes relative to the DST 
    raster scan nearest in time. 
    The positions of the two regions studied in detail 
    in this paper (regions
    A and D) are marked with crosses.
    \label{fig:regions}
  }
\end{figure*}

Another SOHO instrument, the Extreme ultraviolet Imaging Telescope
\citep[EIT;][]{Delaboudiniere-etal:95} provided synoptic series of full disk
images centered on 17.1, 19.5, 28.4 and 30.4~nm around 13:00 and 19:00 UT,
plus a series of full disk images centered on 19.5~nm acquired with a 12
minutes cadence during the routine \textit{CME watch} program.
Fig.~\ref{fig:regions} shows a few images extracted from the EIT~19.5
sequence, showing the development of the flare. 

\begin{figure}[h!tb]
  \centering
  \includegraphics%
  [width=1.0\linewidth]%
  {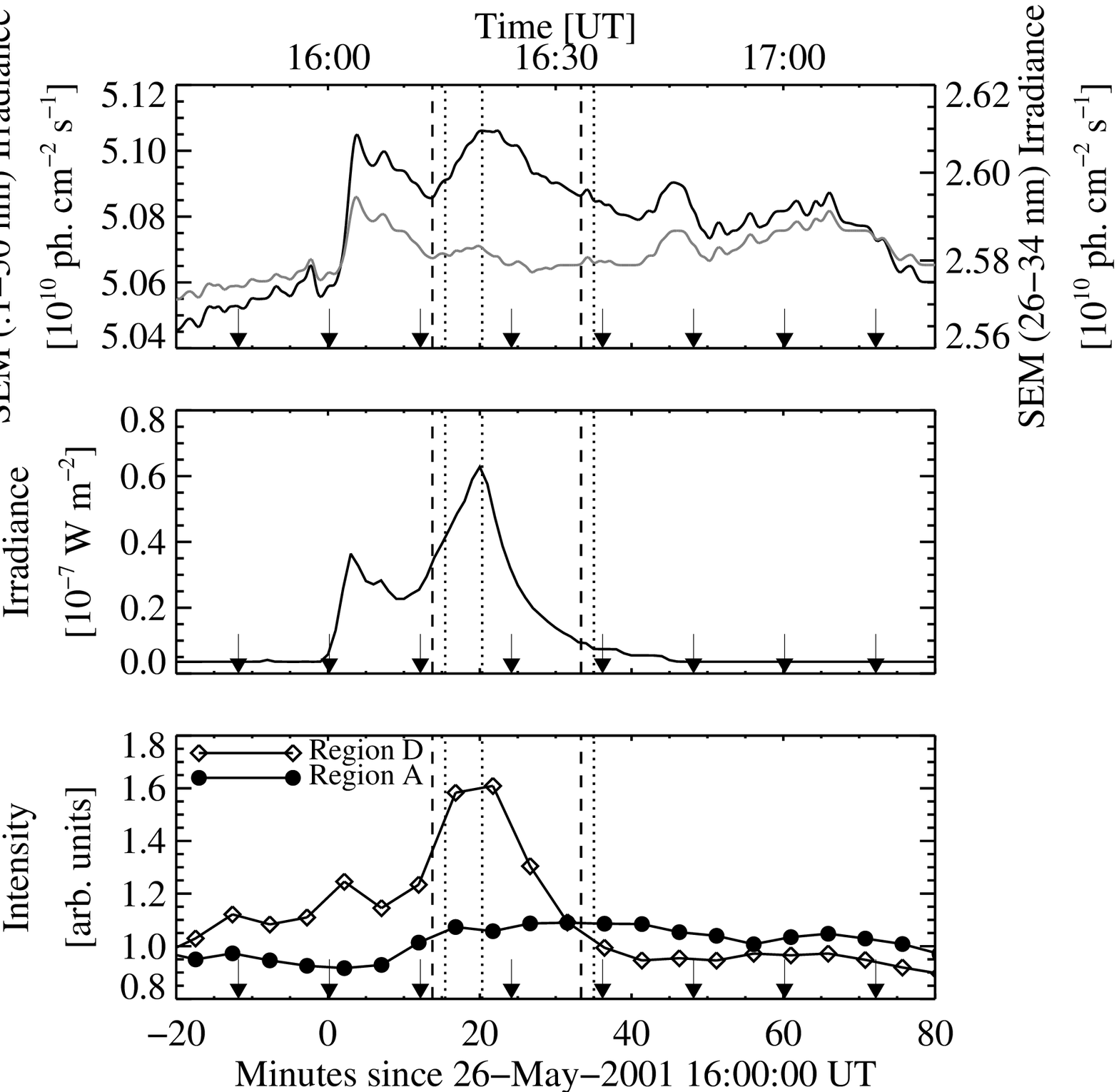}
  \caption{%
    Light curves in the EUV and soft-X bands (1-minute resolution), before and
    during the C-class flare studied in this paper, compared with optical
    measurements. 
    \textit{Top panel:} 
    EUV Irradiance  at 1~AU, from
    SOHO/\SEM\ (black: $\lambda<50$~nm; grey: 
    $26<\lambda<34$~nm); 
    \textit{Middle panel:} 
    Soft-X irradiance from GOES-10 (0.05-0.4~nm band);
    \textit{Bottom panel:} 
    \Halpha\ intensity observed at the Dunn Solar
    Telescope (DST) in the two representative regions analyzed in this paper
    (see Fig.~\protect\ref{fig:regions}).  The vertical arrows mark the times
    of the full-disk images in the 19.5~nm band taken by SOHO/EIT during the
    \textit{CME watch} observing program, while the vertical dashed and dotted
    lines mark the times of the spectra of regions A and D, respectively.
    \label{fig:hxr}
  }
\end{figure}

Finally, the EUV flux during the flare (see Figure~\ref{fig:hxr}) was observed
by the Solar EUV Monitor (\SEM) instrument aboard SOHO
\citep{Hovestadt-etal:95}.  The CELIAS/\SEM\ instrument provides calibrated
total photon counts in the range $\lambda<50$~nm\ (zero-th order, \SEM[0]),
and in the range $26<\lambda<34$~nm\ (first order, \SEM[1]), at 1~AU.

The flare dynamics has been discussed in Paper~I, where it has been shown that
a large area of the flaring region is affected by chromospheric evaporation in
a short time interval (16:02:30 - 16:05:30 UT) during the first impulsive
phase. In the present paper we analyze the He line profiles after 16:13 UT,
when the chromospheric evaporation is almost finished, in two distinct
locations, which we term regions ``A'' and ``D'' (size of $4\arcsec\times
2\arcsec$ each).  These two regions are marked in 
Fig.~\ref{fig:regions}

From what we obtained in Paper~I, region~A did not undergo any evaporation
process, at least within our limit of detectability, while region D is in the
area where we measured downflows at chromospheric levels and upflows at
Transition Region (TR) and coronal levels. Moreover, while in region~A the
chromospheric radiance is almost constant after 16:13 UT, in region~D it
approximately follows the EUV light curve (see Figure~\ref{fig:hxr}) with a
{secondary maximum at about 16:20 UT, showing that also this small
  flare is a multiple event.}
A similar behaviour is 
also detected for the TR and coronal lines with temperatures up to $10^6$~K.

We use region~A as a control region where only small changes are detected in
our observed features.  We construct (Sec.~\ref{sec:models:atmo})
semiempirical models of the flaring atmosphere for this region at two times
when the \SEM\ EUV irradiance is nearly the same: 16:14 and 16:33 UT (dashed
lines in the SOHO/SEM light curve in the upper panel of Fig.~\ref{fig:hxr}).

For region~D the models are constructed at 16:15, 16:20, and 16:35~UT (dotted
lines).  These times will be also referred to as $t_1$, $t_2$, and $t_3$,
respectively.
Time $t_2$ corresponds to a peak of the total \SEM\ irradiance, while at $t_1$
and $t_3$ the irradiances are almost equal. Therefore, the choice of these
regions and times allows us to check whether and how the \ion{He}{1} lines are
influenced by the EUV coronal back-radiation, and/or by the atmosphere where
they form.  This point will be discussed more in detail in
Sec.~\ref{sec:models:jcor:spectrum}.

\begin{figure*}[h!tb]
  \centering
  \includegraphics%
  [angle=90,width=1.0\linewidth]%
  {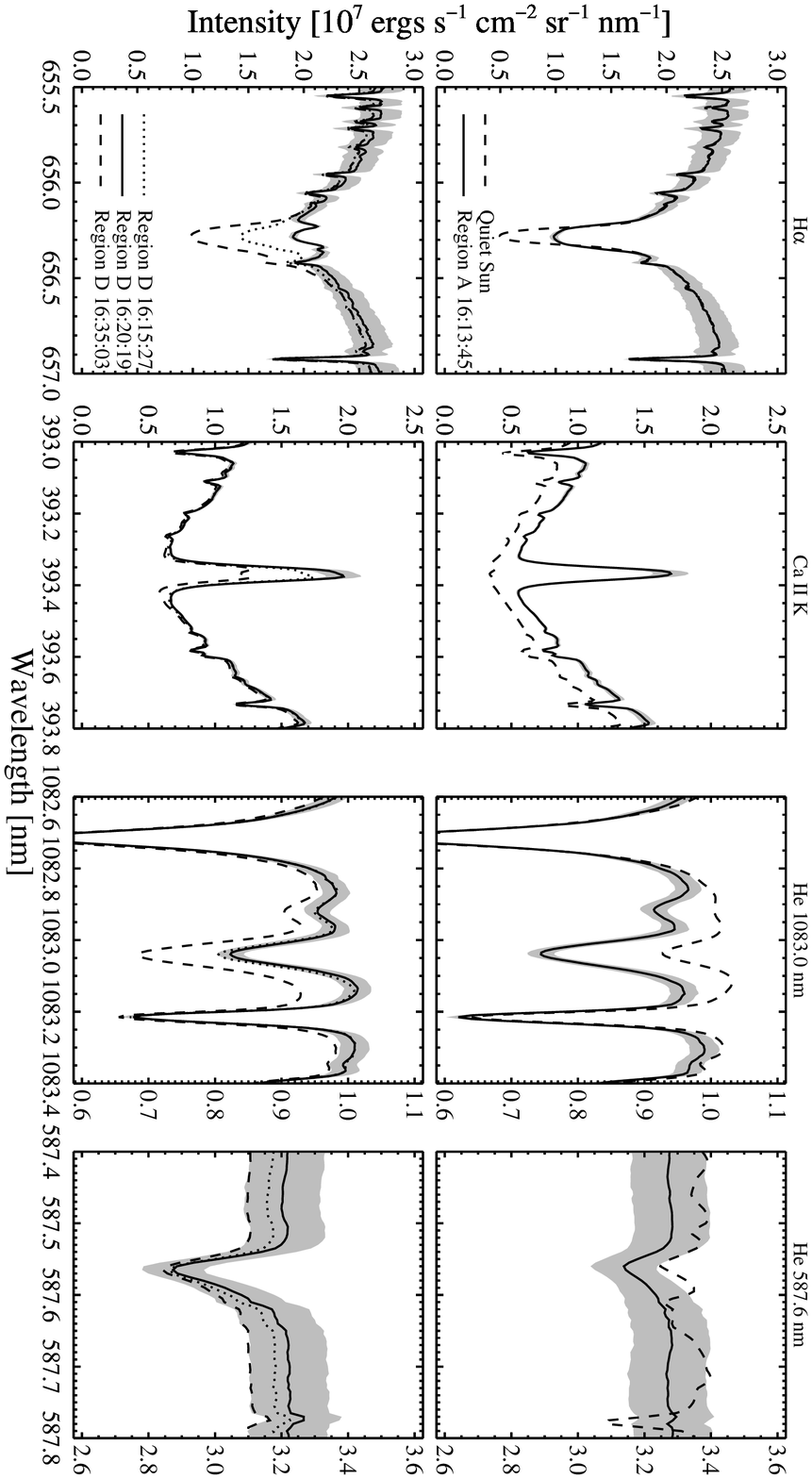}
  \caption{%
    Profiles for the lines \Halpha, \CaK, \ion{He}{1}~1083.0 and
    587.6. 
    \textit{Top panels:} %
    Observations in a quiescent, reference region 
    (dashed lines) and 
    in region~A at 16:14~UT 
    (solid lines). 
    \textit{Bottom panels:} %
    Observations in region~D at 
    16:15 (dotted lines), 
    16:20 (solid lines), and
    16:35~UT (dashed lines).
    Error bands for the profiles in region~A at 16:14~UT and
      region~D at 16:20~UT are also shown.
    Note: telluric lines have been removed from the
    \ion{He}{1}~587.6 profiles, except in the case of the reference region.
    \label{fig:obs_profiles}
  }
\end{figure*}

The two sets of optical line profiles are shown in
Fig.~\ref{fig:obs_profiles}, with the exception of the \ion{Na}{1} doublet
that did not change during the flare in the two considered regions.  For
region~A, since the profiles remain practically unchanged, only the profiles
observed at 16:14 are shown and compared with profiles taken in a nearby,
relatively quiescent region, prior to the flare, which have been used to
calibrate the spectra in absolute units.

The noise of individual spectra is sufficiently small to be negligible
compared to the pixel-to-pixel variability within the target regions which,
therefore, determines the uncertainty on the mean profiles.  The uncertainties
as function of wavelength on the mean profiles in region~A at 16:14~UT and in
region~D at 16:20~UT are indicated in
Fig.~\ref{fig:obs_profiles} as a grey band.  In practice, those bands are
obtained, at each wavelength, from the r.m.s.\ of the $N$ profiles within the
target regions, divided by $\sqrt{N}$.

It is important to realize that for the two relatively weak \ion{He}{1} lines,
the uncertainty in the central line depth (the quantity which carries most
information about the chromosphere) is also due to the variability of the
underlying photosphere.  This fact is most evident in the continuum bands of
the D$_3$ line at 58.76~nm.  By examining the variability of the profiles
normalized to the continuum, we estimate that for the latter line the
photospheric variability contributes by about 50\% to the variability of the
central line intensity%
\footnote{%
  We adopt here, as commonly done in Astrophysics, the term ``specific
  intensity'' or simply ``intensity'' in place of the more formally correct
  ``spectral radiance'' (SI units: W sr$^{-1}$ m$^{-2}$ nm$^{-1}$) and
  indicate it with the symbol $I_\lambda$; with ``radiance'' it is then meant
  an intensity integrated over a wavelength band or a line profile: $\int
  I_\lambda\; \mathrm{d}\lambda$ (SI units: W sr$^{-1}$ m$^{-2}$).  Finally,
  ``irradiance'' is the spectral and angular integral of an intensity: $\int
  I_\lambda\; \mathrm{d}\lambda\:\mathrm{d}\Omega$ (SI units: W m$^{-2}$).%
}.
In any case, the uncertainties shown in Fig.~\ref{fig:obs_profiles} for the
two optical \ion{He}{1} lines (of the order of 3\%), are to be taken as upper
limits for the uncertainties due to the variability of the chromosphere only.

In the case of the EUV \ion{He}{1} lines, the width of the instrumental
response of the CDS spectrograph prevents a detailed comparison of computed
and observed line profiles.  We thus considered only the radiances. For these
lines, the main source of error is the uncertainty in the absolute radiometric
calibration, estimated to be of the order of 30\%.

\section{Coronal back-radiation}\label{sec:models:jcor}

In Paper~II the average EUV illumination over the target area in active region
AR~9468, integrated in the $\lambda<50$~nm range, was
computed using the spatial information provided by the EIT
synoptic images to estimate the contribution to the full-disk \SEM[0]
0.1--50~nm irradiance due to the target region.  This value, \Iref$=1.2\pm 0.2
\times 10^{15}$~photons s$^{-1}$ cm$^{-2}$ sr$^{-1}$, is considered here as
the reference value.
We recall that the emission
from the He II lines has been subtracted from \Iref\ because these lines
are taken into account in a self-consistent way in the model computations.

Following the same approach, and using the EIT~19.5 images as proxies, we
estimated that, in the hour before the flare, the EUV radiation in regions~A
and~D was, in average, about 2.2 and 1.6 times higher than \Iref.  Since there
are variations of the order of 30\% in a $3\times 3$ pixel box ($\approx
16\arcsec\times 16\arcsec$) around the center of each region, the standard
deviation of the \emph{mean} value in the same box would be 3 times smaller.
We however prefer to use the 3-$\sigma$ value on the mean value quoted above,
as it is large enough to encompass other possible sources of error (for
instance those introduced by the use of EIT~19.5 images as proxy for the full
1-50~nm range).  Thus, we estimated the EUV illumination over regions~A and~D
to be, respectively, about $2.6\times 10^{15}$ and $1.9\times 10^{15}$ photons
s${^{-1}}$ cm$^{-2}$ sr$^{-1}$ (Table~\ref{tab:tot_rads}).

The implicit assumption in such an approach, as explained in Paper~II, is that
the spectral distribution of the EUV radiation over the active region does not
change significantly.  For the observations taken during the flare, however,
this assumption may not be valid.  Thus, we adopted an altogether different
approach for estimating the EUV illumination at the times of our observations.

From the EIT sequence of images we can estimate that the only event that can
cause the increase in EUV irradiance shown in the topmost panel of
Fig.~\ref{fig:hxr}, is the C-class flare we observed.  From the
\textit{increase} in \SEM\ irradiance, $\Delta f$, and from the knowledge of
the angular area affected by the flare, $\Delta\Omega$, we can then estimate
the \textit{increase} in radiance in the flaring region, $\Delta I = \Delta
f/\Delta\Omega$.  The total radiance in the region is then obtained by adding
to $\Delta I$ the estimates of the pre-flare average radiances given above.

For a first, rough estimate, we can start from a visual determination, from
Fig.~\ref{fig:hxr}, of the excess irradiance in the \SEM[0] measurements due
to that event: $\Delta f \approx 2$ to $4 \times 10^{8}$ photons s$^{-1}$
cm$^{-2}$ in the first 60 minutes of the event.  On the other hand, from both
EIT and CDS observations, we estimate flare areas to be of the order of
$\Delta\Omega \approx 1500$ to $2000 (\arcsec)^2$ in the same time interval .
Consequently, we expect radiances in the flaring regions to increase by
something of the order of $\Delta I \approx 6$ to $8 \times 10^{15}$ photons
s$^{-1}$ cm$^{-2}$ sr$^{-1}$.

The inaccuracy of this estimate is mainly due to the approximate definition of
the flaring area: it can vary by as much as a factor 2 or 3, depending on
which spectral signature is considered.  For a more precise determination of
the excess irradiances in regions~A and~D,
we refined the method by using the information given by the variation in time
(light curves) \emph{above the pre-flare mean value} of both the SEM
irradiance and of the observed spectral signatures in our data set.  More in
detail, we proceeded as follows:

\textit{Step \#1:} %
We created a set of reference images, both from CDS and from EIT, by averaging
the (co-aligned) images in the 40 minutes preceding the onset of the event.
The variability of these images around the average is of the order of
0.1--0.2\% (standard deviation), or even less.
\label{step:ref_maps}

\begin{figure*}[h!tb]
  \centering
  \includegraphics%
  [angle=90,width=1.0\linewidth]%
  {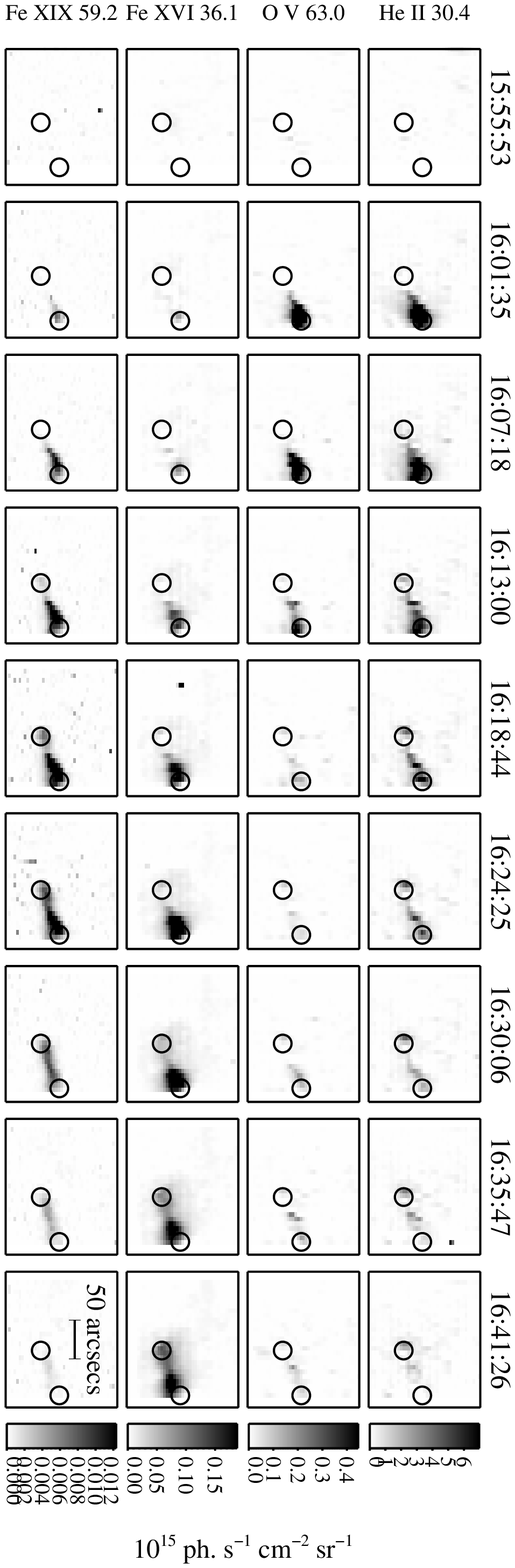}
  \caption{%
    Sequences of CDS/NIS \emph{residual} spectroheliograms showing, for
    about 40 minutes, the evolution of the event from its onset (at about
    16:01~UT). The images obtained from the profiles of the
    \ion{He}{1}~58.4~nm are very similar to those of the \ion{He}{2}~30.4~nm
    line, and thus are not shown here.  Circles mark the two regions of
    interest (A and D: left-most and right-most circles, respectively).
    \label{fig:diff_maps}
  }
\end{figure*}

\textit{Step \#2:} %
We obtained maps of ``residual'' or ``excess'' radiance in each spectral
feature during the flare, $\Delta I_{ij}^k(t_h)$, by subtracting the reference
pre-flare images (taking into account solar rotation) for each time $t_h$, and
for each pixel $(i,j)$.  Multiplying by the pixel angular size,
$\mathrm{d}\omega^k$, we then obtain the contribution of each line and of each
pixel to the EUV irradiance, $\Delta f_{ij}^k(t_h)=\Delta
I_{ij}^k(t_h)\;\mathrm{d}\omega^k$.  The result is shown in
Fig.~\ref{fig:diff_maps} for CDS images only.  A similar procedure has been
applied to the EIT sequence as well.  More details on the analysis of the
individual CDS rasters can be found in Paper~I.  Here we only mention that the
rasters were taken by moving the slit from W to E --- right to left, in the
images of Fig.~\ref{fig:diff_maps}, where the beginning of each raster is
indicated at the top of the corresponding column.
\label{step:diff_maps}
%

\begin{figure}[h!bt]
  \centering
  \includegraphics%
  [width=1.0\linewidth]%
  {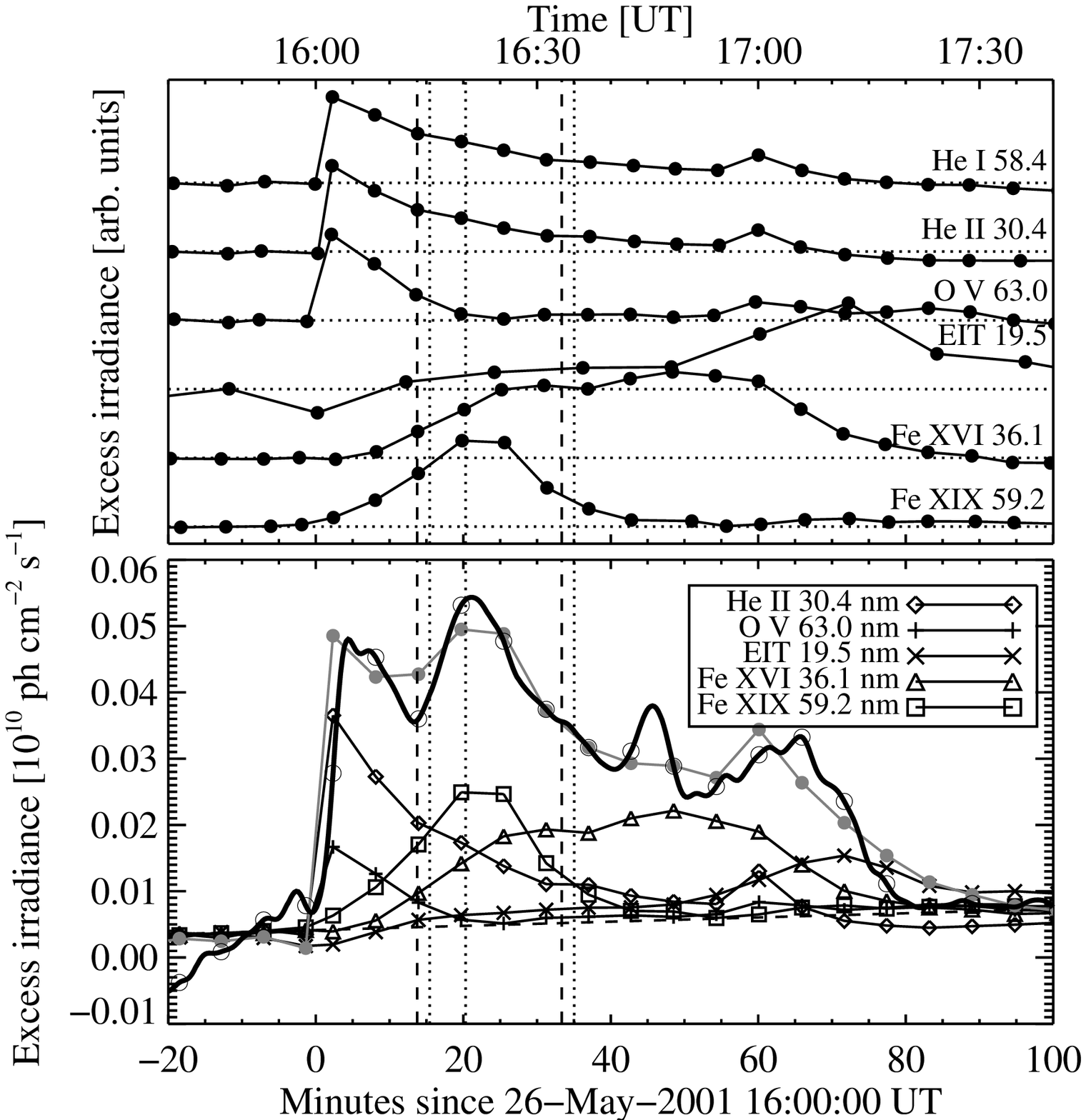} 
  \caption{%
    \textit{Top panel:} %
    Light curves of residual irradiances in CDS lines; the
    EIT enhancement over the flaring region is also shown. %
    \textit{Bottom panel:} %
    Light curve of excess \SEM[0] irradiance during the flare (black, thick
    line,                  and empty circles), compared with the fit (grey line,
                     and filled circles) obtained by combining the CDS
    and EIT light curves. %
    The vertical dashed and dotted lines mark the times of the spectra taken
    in regions A and D, as in Fig.~\protect\ref{fig:hxr}.
    \label{fig:lc_reg}
  }
\end{figure}

\textit{Step \#3:} %
We computed the light curves of the \emph{excess} irradiance for each spectral
feature by summing the contributions $\Delta f_{ij}^k(t_h)$.  Each curve
represents the contribution of the considered feature to the EUV excess
irradiance
during the flare, under the assumption that no other significant event is
occurring on the solar surface outside the CDS FOV.  We verified the
correcteness of that assumption by inspecting the full-disk EIT 19.5 images
during the event. The resulting light curves are shown in the top panel of
Fig.~\ref{fig:lc_reg}.  The set of CDS and EIT curves cover a broad range of
temperatures (from $\log T < 5.0$ for \ion{He}{1} and \ion{He}{2}, to $\log T
\sim 6.9$ for \ion{Fe}{19}), and can thus be regarded as a good set of proxies
for the temporal and spatial evolution of the various temperature components
of the flaring plasma.
  \label{step:light_curves}
  %

\textit{Step \#4:} %
We found a linear combination of those light curves matching reasonably well
the evolution of \SEM[0] measurements during the flare.  More in detail, we
modeled the total \emph{excess} irradiance at each time $t_h$, $\Delta
f(t_h)$, with a linear combination of the \emph{excess} irradiance of each
line $k$, in the form:
\begin{equation}
  \label{eq:sem_fit}
  \Delta f(t_h) \approx a_0 + a_1 t_h + 
  \sum_k b_k \sum_{ij}\Delta f_{ij}^k(t_h)
  \; .
\end{equation}
The term $a_1 t_h$, that accounts for possible overall changes in the active
region structure (hence: of the overall EUV emission) in the course of the
event, is actually rather small: To reduce the number of free parameters, we
set this term to zero.

The two resonance helium lines, \ion{He}{1}~58.4 and \ion{He}{2}~30.4, exhibit
an almost identical behaviour: only the latter has been taken into account in
the linear regression.

The comparison of the thus modelled EUV SEM irradiance (grey line in
Fig.~\ref{fig:lc_reg}) with the actual measured signal (smoothed to account
for the 5-minutes duration of the CDS rasters) is shown in the bottom panel of
Fig.~\ref{fig:lc_reg}, for the first $\approx 120$ minutes of the event.  We
stress that $b_k \sum_{ij}\Delta f_{ij}^k$ is \emph{not} the contribution of
the individual line $k$ to the EUV irradiance, but is instead the contribution
of the \emph{group} of lines of similar temperatures, for which line $k$ is a
proxy.

Beyond the mere application of the mathematical linear regression procedure,
we think useful to remark here how the different parts of the irradiance \SEM\
light curve are due to different temperature contributions of the flare
plasma.  More specifically, the first peak, at the time resolution of the CDS
rasters, is modelled essentially by the cooler temperature lines (proxies:
\ion{He}{2} and \ion{O}{5}), while the second peak (at $\sim$16:20~UT) is due
mostly to the hottest flaring plasma (proxy: \ion{Fe}{19}), with a tail
modelled by progressively cooler components (\ion{Fe}{16} and then EIT~19.5).
These finding, based on the modelling of the \SEM[0] light curve, are
qualitatively consistent with the characteristics of the other irradiance
curves shown in Fig.~\ref{fig:hxr}: the second peak is indeed absent in
\SEM[1], while is more pronounced in the GOES signal.
\label{step:lin_regression}

\begin{figure*}[h!bt]
  \centering
  \includegraphics%
  [angle=90,width=1.0\linewidth]%
  {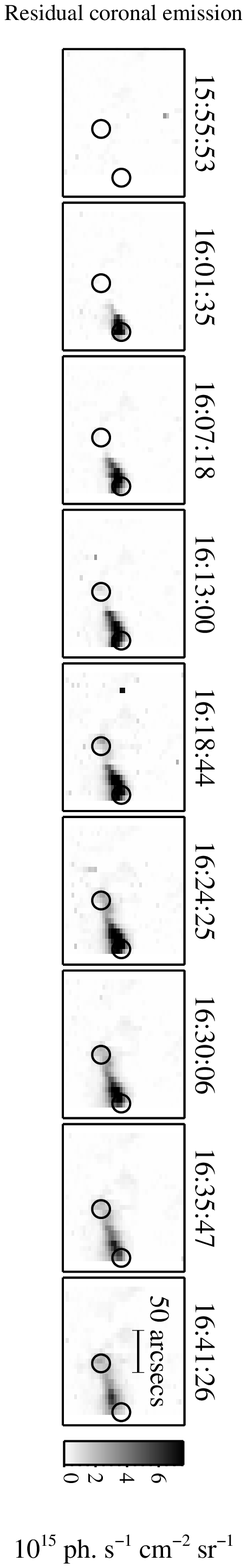}
  \caption{%
    Sequences of images showing the variation of the \emph{enhancement} of EUV
    illumination in the $\lambda<50$~nm band, in the flaring area.
    Radiance maps shown with the same format as in
    Fig.~\protect\ref{fig:diff_maps}.
    \label{fig:rad_maps}
  }
\end{figure*}

\textit{Step \#5:} %
The coefficients of the linear combination, $b_k$, were then applied to the
corresponding residual images on a pixel-by-pixel basis, producing a map of
the contribution to the total \emph{excess} EUV irradiance in the SEM spectral
range ($\lambda<50$~nm) as function of time during the flare.  The
contribution due to transition region and coronal lines (henceforth, for
brevity: ``the coronal component''), needed in the model calculations, is
obtained by subtracting the contribution of the proxy \ion{He}{2}~30.4~nm,
that will be computed self-consistently in the models.
Thus, the contribution of each pixel to the EUV coronal irradiance, divided by
the pixel's angular size, yields an estimate of the spatial variation of
coronal radiance \emph{enhancement}, in the band $\lambda<50$~nm, over the
flare area.
The result of such a procedure is shown in Fig.~\ref{fig:rad_maps}.

If we make the reasonable assumption that the EUV emission during the flare is
optically thin, the maps of EUV \emph{excess} radiance of
Fig.~\ref{fig:rad_maps} are also maps of EUV back-radiation illuminating the
chromosphere.  In Sec.~\ref{sec:models:jcor:spectrum} we will discuss more in
detail this assumption.
\label{step:tot_maps}
%

\textit{Step \#6:} %
Finally, by adding the pre-flare, average value of radiance estimated in the
two target regions, A and~D, we obtained an estimate of the value of EUV
radiance of coronal origin in the band $\lambda<50$~nm, \Icor, as function of
time over the two target regions,
and in particular at the times of the optical observations of
Fig.~\ref{fig:obs_profiles} (Table~\ref{tab:tot_rads}).
\label{step:tot_rads}
%

\begin{table}[h!tb]
  \begin{center}
    \begin{tabular}{lcccc}
      \tableline
      \tableline
      Region & Time      & \multicolumn{3}{c}{Coronal radiance} \\
             &  (UT)     & Excess & Total & Total/\Iref \\
      \tableline
      A      & pre-flare &  ---   & $2.6$ & 2.2         \\ 
             & 16:14     & $0.38$ & $3.0$ & 2.5         \\ 
             & 16:33     & $1.73$ & $4.4$ & 3.6         \\ 
      \tableline
      D      & pre-flare &  ---   & $1.9$ & 1.6         \\ 
             & 16:16     & $5.48$ & $7.3$ & 6.2         \\ 
             & 16:20     & $7.38$ & $9.2$ & 7.7         \\ 
             & 16:35     & $5.25$ & $7.1$ & 6.0         \\ 
      \tableline
    \end{tabular}
  \end{center}
  \caption{%
    Coronal radiances for regions~A and~D, in units of  $10^{15}$
    photons s${^{-1}}$ cm$^{-2}$ sr$^{-1}$.
    The value $\Iref=1.2\times 10^{15}$ is the reference (average
    pre-flare) value for the
    target area, estimated in Paper~II.
    \label{tab:tot_rads}
  }
\end{table}

The (statistical) uncertainties in the excess number of photons are of the
order of $\approx 0.4\times 10^{15}$ (region A) and $\approx 2\times 10^{15}$
(region D) photons s$^{-1}$ cm$^{-2}$ sr$^{-1}$, estimated 
as the 3-$\sigma$ variation of the illumination around the target regions.

We have also explored the systematic variations induced by changing the most
relevant parameters that could affect the result of the procedure described
above (for example: using a constant instead of a linear background in the
regression calculations of Step \#4); 
the resulting overall variations are smaller than the statistical uncertainty
at the 3-$\sigma$ level quoted above.  

Together with the uncertainty in the pre-flare radiance enhancements estimated
earlier in this section ($\approx 30$\%), and with the uncertainty of \Iref\
estimated in Paper~II ($0.2\times 10^{15}$, or about 17\%), we thus obtain an
overall uncertainty in the total coronal radiance of the order of 25\% to 35\%
in both regions A and D.

\subsection{The spectral and angular distribution of the coronal
  back-radiation}\label{sec:models:jcor:spectrum}
  
In Paper~II we estimated the spectral distribution of the coronal
photoionizing EUV radiation by taking a reference spectral distribution, and
multiplying that distribution by a factor (close to unity)
such that the radiance in the range $\lambda<50$~nm equals $\Iref=1.2\times
10^{15}$~photons s${^{-1}}$ cm$^{-2}$ sr$^{-1}$.
    
In that paper we adopted as our reference spectral distribution the one for a
``typical'' active region computed with the spectral code and atomic database
CHIANTI version~4 \citep{Young-etal:03}.  The CHIANTI calculations require a
description of the density and temperature distribution in the atmosphere via
the Differential Emission Measure (\DEM), defined as
$\NH\Ne(\mathrm{d}T/\mathrm{d}h)^{-1}$, where $h$ is a coordinate along the
line of sight; we adopted the table in file \texttt{active\_region.dem}
provided with the package, which is based on data from
\cite{Vernazza-Reeves:78}.  Furthermore, we adopted a ``hybrid'' set of
abundances \citep[intermediate between photospheric and coronal elemental
mixtures:][]{Fludra-Schmelz:99}, and a constant pressure of $\Pe/k = 3\times
10^{15}\;\mathrm{cm}^{-3}\;\mathrm{K}$.
We verified that the use of the more
recent version~5 of the same code \citep{Landi-etal:06} did not alter
significantly that reference spectral
distribution (Fig.~1 of Paper~II).

In the case of the flaring atmosphere we are considering, however, it is
possible that significant changes in the spectral distribution may occurr.
We tested this possibility by considering the case where departures from the
``standard'' spectral distribution (the pre-flare AR EUV spectrum, in fact)
could be more pronounced: the emission above region~D at 16:20~UT, where we
observe a peak of total EUV emission (Table~\ref{tab:tot_rads}), as well as a
peak in the hot \ion{Fe}{19} line (Fig.~\ref{fig:lc_reg}).

\begin{figure*}[h!tb]
  \centering
  \includegraphics%
  [bb=0 41 267 798,angle=90,width=1.0\linewidth]%
  {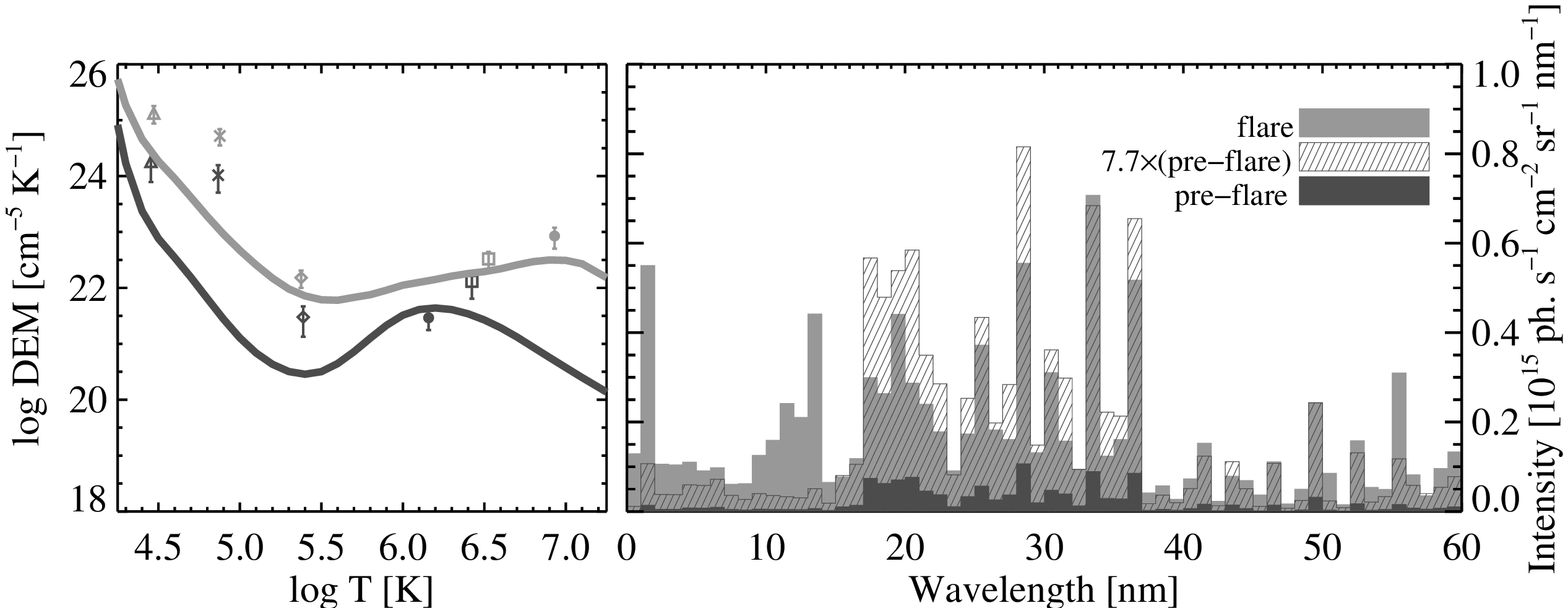}
  \caption{%
    \textit{Left:} %
    \DEM\ 
    distributions for the ``reference'' AR atmosphere (black line) and the
    flaring region D at 16:20~UT (grey line, yellow in the on-line version).
    The observed intensities (multiplied by the factor
    $\DEM(\Teff)/I_\mathrm{calc}$) are also shown for the various CDS lines:
    \ion{He}{1}~58.4~nm (triangles), \ion{He}{2}~30.4~nm ($\times$ symbols),
    \ion{O}{5} 62.9~nm (diamonds), \ion{Fe}{16}~36.1~nm (squares), and the
    blend at 59.2~nm (filled circles).  
    \textit{Right:} %
    Spectral distributions from the above \DEM s,
    excluding all \ion{He}{1} and \ion{He}{2} lines and continua, compared
    with the ``standard'' AR emission times the multiplicative factor of
    $\approx 7.7$ given in Table~\protect\ref{tab:tot_rads}
    (dashed histogram).
    \label{fig:dem_spec}
  }
\end{figure*}

To do so, we found a \DEM\ producing the measured total number of photons in
the range $\lambda<50$~nm ($9.2\times 10^{15}$ photons s${^{-1}}$ cm$^{-2}$
sr$^{-1}$), and giving at the same time a reasonable match to the
transition-region and coronal lines observed by CDS: \ion{O}{5} 62.9~nm,
\ion{Fe}{16}~36.1~nm, and the blend \ion{Fe}{12}~59.26+\ion{Fe}{19}~59.22~nm.
The optically thick helium lines have not been considered in this analysis,
but are nonetheless shown in the plots summarizing the results of this test
(Fig.~\ref{fig:dem_spec}).  

In the calculations, we assumed a higher pressure, $\Pe/k= 3\times
10^{16}\;\mathrm{cm}^{-3}\;\mathrm{K}$, than in the pre-flare atmosphere.
This value of the pressure corresponds, for instance, to $\Ne\approx 10^{11}$
and $\approx 10^{10}$ at $\log T=5.5$ and $6.5$ respectively.  These densities
are generally consistent with flare densities found in the
  literature \citep[e.~g.:][]{Dere-Cook:79}, although lower densities have
sometimes been reported \citep{Cook-etal:94}.  Aside of the overall scale
factorized by the \DEM, the specific value for $\Pe/k$ affects only the
relative radiances of the density-dependent lines, and thus it is of little
importance for the purpose of estimating the \emph{global} flare spectral
distribution for $\lambda<50$~nm.

Such a \DEM\ is compared in the left-hand panel of
Fig.~\ref{fig:dem_spec} with the \DEM\ used to compute the pre-flare,
reference spectrum.  The corresponding observed line radiances are shown in
that panel multiplied by the factor $\DEM(\Teff)/I_\mathrm{calc}$ \citep[as
in, e.~g., ][]{DelZanna-etal:02}; here $\Teff$ is the temperature of peak line
emission.  The error bars in the case of the data taken during the flare
reflect the statistical variability in the 9 pixels around and above the
target (D) region.  Thus, we do not show the theoretical uncertainties or the
calibration uncertainties both on the absolute intensities and on the CDS
NIS1/NIS2 relative radiances (i.~e.: the radiance of the \ion{Fe}{16} line
relative to all the others).

The \DEM-derived pre-flare and the flare
spectra are compared in the right-hand panel
of the same figure (dark and light grey histograms, respectively).  We also
show the pre-flare ``standard'' spectrum multiplied by the factor required to
obtain the same total number of photons below 50~nm as in the flare spectrum
(dashed histogram). The largest differences between these two spectra occur in
the region below 15~nm, where the number of photons in the modelled flare
spectrum is about a factor 4 higher than in the AR scaled spectrum.
Conversely, the number of photons in the range 15--50.4~nm is about 20\%
smaller than in the AR scaled spectrum.

On the other hand, for both distributions the bulk of the photoionizing
photons is in the range $15<\lambda<50$~nm (about 3/4 of the total).  In fact,
this is the main reason why, as discussed in Paper~II, the shorter wavelength
part of the spectral distribution has relatively little effect on the
photoionization of helium: We found that only the \ion{He}{1}~1083 line shows
small changes due to the different penetration depths in the chromosphere of
the EUV photons of different energies.  We will further discuss this point in
Sec.~\ref{sec:models:atmo:jcor}.

In conclusion, we expect, and indeed find (see
Sec.~\ref{sec:models:atmo:jcor}), that for our calculations it is not
necessary to take into account the details of the distribution of photon
energies: after all, we are studying a quite small flare.  We thus simply
scaled the ``standard'' spectral distribution by the appropriate factors
listed in the last column of Table~\ref{tab:tot_rads}.

Concerning the problem of the angular distribution of the EUV photoionizing
photons, we have already showed in Paper~II that, despite their relatively
featureless spatial distribution across the target AR prior to the flare, at
the chromospheric depths the details of the angular distribution are
relatively unimportant compared to the overall spectral distribution and, most
importantly, to the total number of impinging photons.

In the case considered in this paper, on the other hand, during the flare
the EUV emission is clearly concentrated in smaller areas
(Figs.~\ref{fig:diff_maps} and~\ref{fig:rad_maps}).  It is therefore
reasonable to assume that the main sources of ionizing radiation are
concentrated directly above, or very near the target regions.
We verified this assumption using the control region (region A): during the
time span we are considering, the helium lines in that region hardly change
(Fig.~\ref{fig:obs_profiles}), despite the large variations of the EUV
radiation around (above) region~D (see Table~\ref{tab:tot_rads}), which is
roughly 40~Mm away.  So, it is clear that the flaring chromosphere in
region~A, and \textit{a fortiori} in region~D, is sensitive only to the EUV
radiation coming from regions significantly closer than 40~Mm $\approx
50\arcsec$, or, in other terms, is sensitive to the radiation coming at most
from a few EIT or CDS pixels away.

\section{Chromospheric modeling}\label{sec:models:atmo}

Once the coronal back-radiation has been determined, semi-empirical models
were constructed to match the observed profiles.  The modeling was done using
the program PANDORA \citep{Avrett-Loeser:84}. Given a $T$ vs.\ $h$
distribution, we solved the non-LTE radiative transfer and the statistical and
hydrostatic equilibrium equations, and self-consistently computed non-LTE
populations for 10 levels of H, 29 of \ion{He}{1}, 15 of \ion{Fe}{1}, 9 of
\ion{C}{1}, 8 of \ion{Si}{1}, \ion{Ca}{1} and \ion{Na}{1}, 6 of \ion{Al}{1}
and \ion{He}{2}, and 7 of \ion{Mg}{1}. In addition, we computed 6 levels of
\ion{Mg}{2}, and 5 of \ion{Ca}{2}. The atomic models we used for H and
\ion{Ca}{2} are described in \cite{Mauas-etal:97} and \cite{Falchi-Mauas:98}.
The models we used for \ion{He}{1} and \ion{He}{2} are described in Paper~II.
More details on the modeling, and on the different assumptions and their
validity can be found in \cite{Falchi-Mauas:98} and in Paper~II.  The adopted
microturbulence, in particular, is the same as in the latter work.

  The modeling was done in a plane-parallel atmosphere.  This is certainly an
  adequate approximation if we consider the much larger horizontal size of the
  target areas (several thousands of km) compared with the thickness of the
  chromosphere under study (a few hundred km).  The effect of a possible
  filamentary structure of the atmosphere is however much more difficult to
  assess.

  The only (qualitative) assessment we can make, from a radiative transfer
  point of view, is that photons could perhaps escape more easily from a set
  of randomly oriented, high-density structures embedded in a lower density
  medium than from a homogeneous atmosphere, especially in the case of high
  density contrast and of a sparse distribution of the filamentary structures.
  Thus, we may expect some modifications in the details of the profiles of the
  more optically thick lines: for instance, a reduction of the depth of
  self-reversed cores, if present.  We may also expect that (effectively)
  optically thin lines should be least affected by atmospheric
  inhomogeneities.

  In the case of more ordered, more closely packed, and thicker structures on
  the other hand, the outcome could be somewhat different, as shown by
  \cite{Gunar-etal:07}, who studied the formation of the hydrogen Lyman lines
  in prominences, modelled as a set of parallel, finite slabs (``threads''),
  each 1000~km wide: each one, that is, as wide as the entire chromosphere we
  are modelling (Figures~\ref{fig:mod_kernel_A} and \ref{fig:mods_kernel_D}).
  We remark, however, that even in such a relatively extreme case, the changes
  are typically not very dramatic, and observable only at high spectral
  resolution.

  In any case, there is little or no information, to our knowledge, on the
  effect of a flare on the chromospheric fine structure.  Thus any more
  precise assessment on the effect of non-uniformity on the emerging profiles
  is virtually impossible at this stage.

In Fig.~\ref{fig:obs_profiles} we have shown the sets of
optical profiles used in the modeling procedure. Since different
features of the lines are sensitive to modifications in different
parts of the chromosphere, the models were constructed following
several steps. First, we fixed the deepest part of the atmosphere
by matching the wings and the K$_1$ minimum of the \CaK\ line, which
are formed in the high photosphere and the temperature minimum
region.

As a second step, we modified the chromosphere and the lowest part of the
transition region until a satisfactory match was found for the profiles of
\Halpha, \CaK, and \ion{He}{1}~1083 and 587.6.  We recall that the profiles of
the first two lines do not depend on the coronal back-radiation field.

The profiles of the two helium lines are formed in two distinct regions: most
of the radiation is originated in the photosphere, which in the quiet Sun
results in a weak absorption line at 1083 nm, and no noticeable line at 587
nm. However, already in the active region studied in Paper~II we found that
there is an important chromospheric contribution. This contribution becomes,
of course, much more important in the flaring atmospheres we are studying
here, and depends not only on the thermal structure of the high chromosphere
and the low transition region, between $10^4$ and $2.5\times 10^4$ K, but also
on the coronal EUV incident radiation.

Finally, both ultraviolet lines, the \ion{He}{1} line at 58.4 nm and
the \ion{He}{2} line at 30.4 nm, depend on the structure of the low
and mid-transition region, from                  $3\times 10^4$  to                  $5\times10^4$  K
for the 58.4 line, and up to                  $1\times 10^5$  K for the 30.4 line. As
a third step, therefore, we modified the structure of this region to
obtain a good match for the line fluxes of these UV lines, measured
with SOHO/CDS. Generally, it was necessary to iterate between steps
2 and 3 before a final model was obtained.

We first computed models for regions A and D, marked in 
Fig.~\ref{fig:regions}, using the coronal radiation determined in
Sec.~\ref{sec:models:jcor} and the same value of the helium abundance
(\AbHe=0.1) we used in Paper~II.  The resulting computed intensity profiles
were convolved with the appropriate instrumental response.

\subsection{Modeling of region A}\label{sec:models:atmo:A}

\begin{figure}[h!tb]
  \centering
  \includegraphics%
  [width=1.0\linewidth]%
  {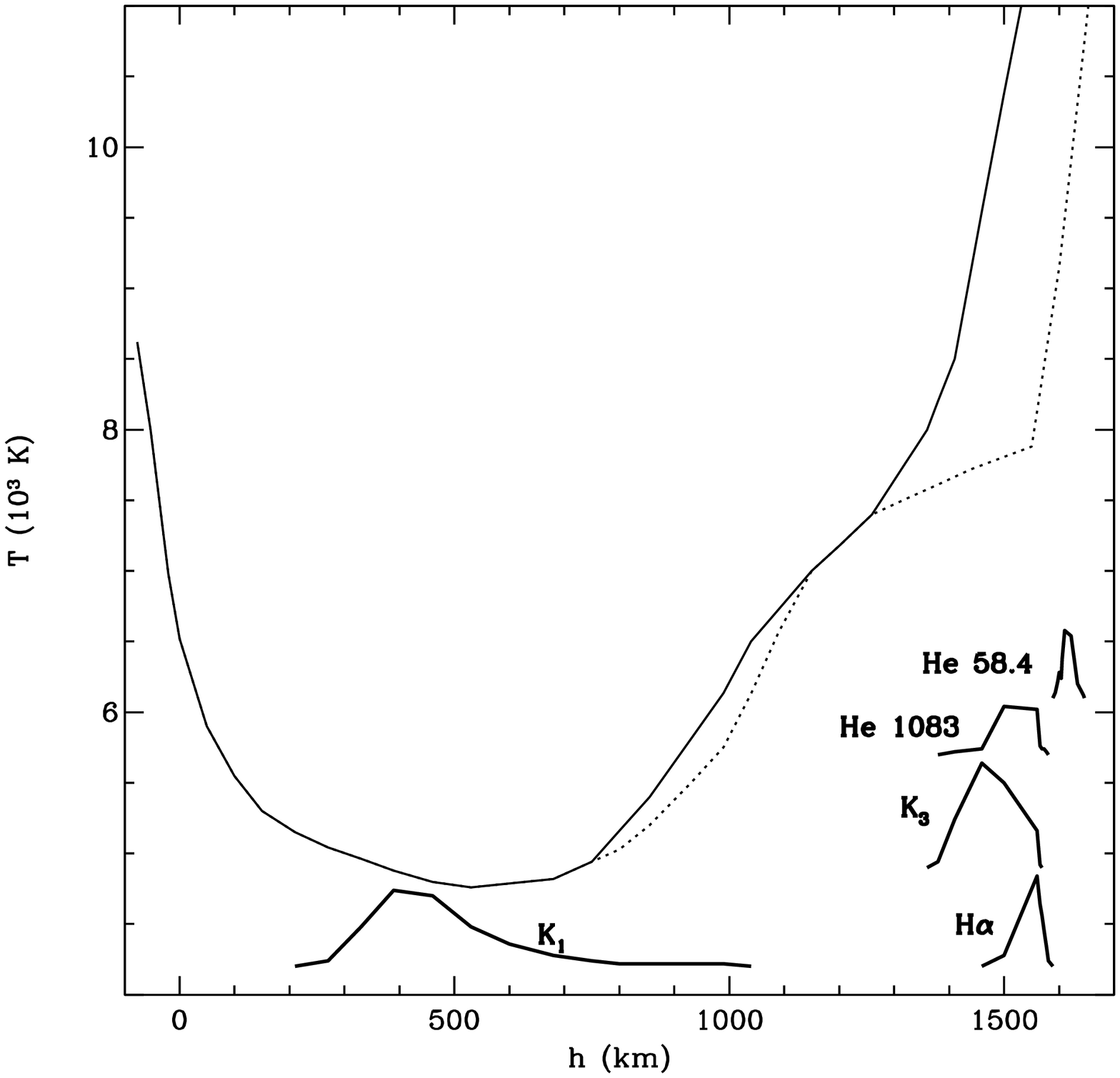}
  \caption{%
    Region A: temperature versus height distribution (solid line) of
    the atmospheric model obtained for \AbHe=0.1. The model for the
    active region studied in Paper II (dotted line) is displayed for
    comparison. 
    Also shown are the
    intensity contribution functions, in arbitrary units
    (solid, thick lines), 
    of the different spectral features we used to build the model.
    The \ion{He}{1}~587.6 contribution function is
    proportional to the one of the infrared \ion{He}{1} line. 
    \label{fig:mod_kernel_A}
  }
\end{figure}

For region A, where the observed lines do not change during the considered
time interval, we obtained the model shown in Fig.~\ref{fig:mod_kernel_A},
which gives a good match between the observed and computed lines.  In the same
figure, we also show, for the line cores, the intensity contribution
functions, defined as
$\mathrm{d}I/\mathrm{d}h = S \exp(-\tau) \chi$, where $I$ is the intensity,
$h$ is the height in the atmosphere, $S$, $\tau$ and $\chi$ are, respectively,
the source function, the optical depth, and the opacity in the line core.

\begin{figure*}[h!tb]
  \centering
  \includegraphics%
  [angle=90,width=1.0\linewidth]%
  {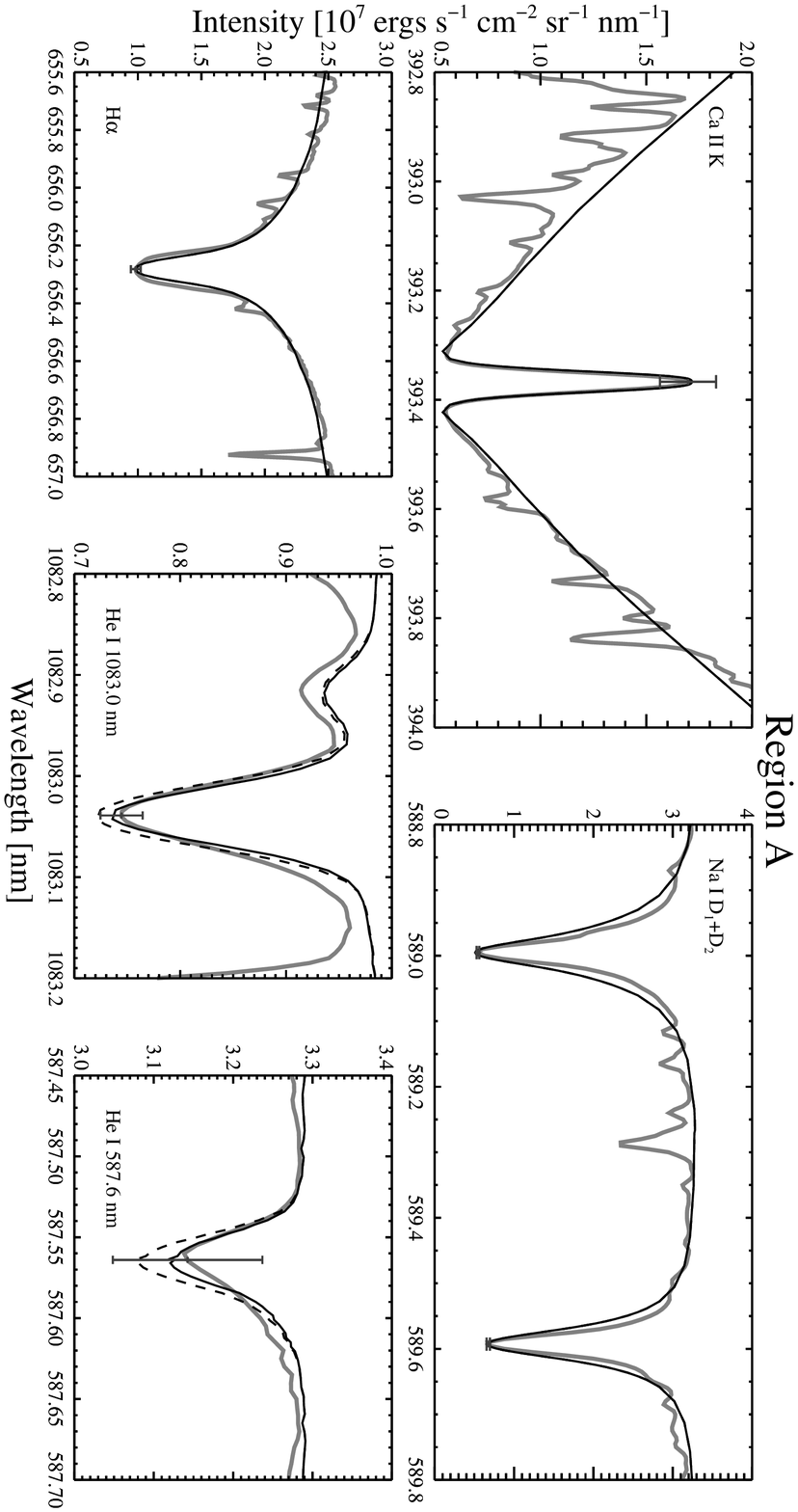}
  \caption{%
    Region~A: comparison between observed (thick grey line) and computed
    profiles (black lines).  The solid and dashed lines
    indicate the profiles computed for the radiation determined at 16:14
    and 16:33~UT, respectively.
    \label{fig:comp_profiles_A}
  }
\end{figure*}

The observed and computed profiles of the lines formed essentially at
chromospheric levels are shown in Fig.~\ref{fig:comp_profiles_A}.  The bars in
the Figure indicate the uncertainty of the observed profiles (see also
Fig.~\ref{fig:obs_profiles}). It can be seen that the agreement found is very
good, well whithin the error bars, in all cases.  In the case of the optical
\ion{He}{1} lines, however, we recall that, as discussed in
Sec.~\ref{sec:obs}, the error bars in the line centers also include the effect
of variations of the underlying photosphere within the target region, and
therefore represent an upper limit to the variation induced by chromospheric
changes.

It can also be seen that only the He lines are affected by the EUV radiation,
and that the computed profiles agree with the observations within the range of
variability in the target region for both values of the coronal radiation
indicated in Table~\ref{tab:tot_rads}.

We have already mentioned in Sec.~\ref{sec:obs} that the UV lines (the
\ion{He}{1} 58.4 and the \ion{He}{2} 30.4) have a large width due essentially
to the instrumental profile: therefore the comparison with the models can be
done only through the radiances, which are shown in Table~\ref{tab:UV_flux}.
It can be seen that it is possible to obtain line radiances in good agreement
with the observed ones using the same atmosphere, despite the variation of the
coronal radiation of about 50\% during the considered time interval (see
Sec.~\ref{sec:models:jcor} and Table~\ref{tab:tot_rads}).

We remark here that the temperature structure in the region of formation of
the \ion{He}{2} 30.4~nm line is not constrained by other lines.  Thus, for the
remainder of the discussion, we will not consider a match of the \ion{He}{2}
\Lalpha\ to be a significant constraint on the atmospheric parameters.

\subsection{Modeling of region~D}\label{sec:models:atmo:D}

For region D, where the chromospheric intensity (Fig.~\ref{fig:hxr})
approximately follows the EUV light curves, we were able to find models
reasonably matching the observations at $t_1$ and $t_3$, but it was not
possible to obtain a model that would simultaneously reproduce all the
profiles observed at $t_2$.
  
In particular, the radiances of the \ion{He}{1} 58.4 line always results
considerably higher than the observed one if we match the profiles of the
other lines, and in particular of \Halpha.  Conversely, it is possible to find
models that reproduce the radiance of the resonance \ion{He}{1} line, but in
this case the other lines, in particular \Halpha, are poorly fitted.  It
should be kept in mind that it is precisely at this time when the regions of
formation of the 58.4 line and of \Halpha\ overlap most, and therefore the
model is best determined (see Sec.~\ref{sec:models:atmo:abund}).

\section{Discussion}\label{sec:disc}

In order to find in region D a model of the flaring atmosphere that matches
all the observed lines during the considered time interval, and in particular
that matches \emph{both} \Halpha\ \emph{and} \ion{He}{1} 58.4, we explored the
two aspects that most likely affect the helium spectrum: the chromospheric
helium abundance, \AbHe\ (Sec.~\ref{sec:models:atmo:abund}), and the spectral
distribution and overall level of EUV coronal back-illumination
(Sec.~\ref{sec:models:atmo:jcor}).

\subsection{The effect of helium abundance}\label{sec:models:atmo:abund}

\begin{figure}[h!tb]
  \centering
  \includegraphics%
  [trim=40 0 0 0,width=1.0\linewidth]%
  {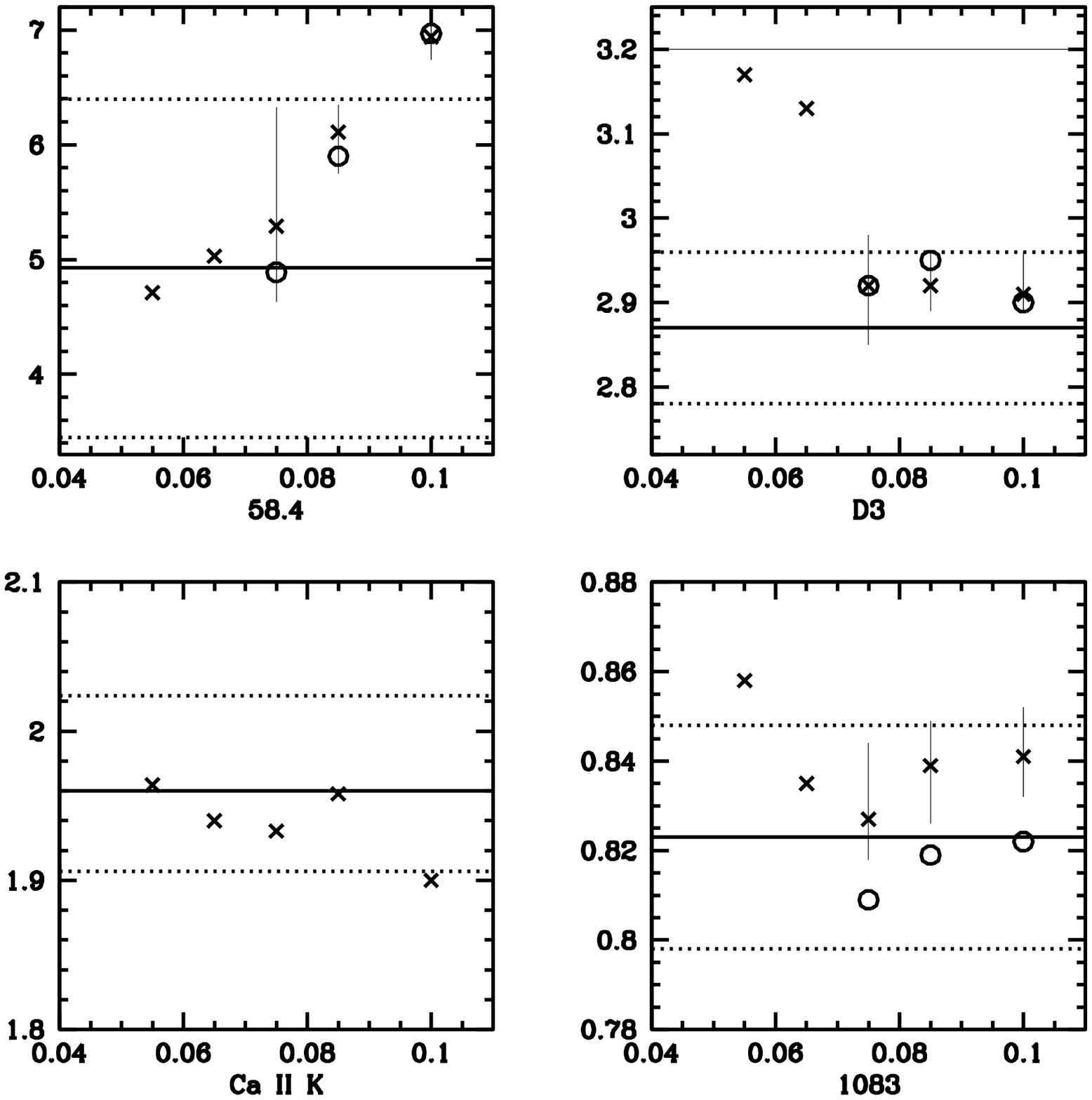}
  \caption{%
    Computed radiance (10$^4$ ergs s$^{-1}$ cm$^{-2}$
      sr$^{-1}$) and central line intensity (10$^7$ ergs s$^{-1}$ cm$^{-2}$
      sr$^{-1}$ nm$^{-1}$) versus \AbHe.  Radiance of the 58.4 line and the
      line center intensity of the D$_3$, 1083 and \CaK\ lines are computed
      for the best-matching models obtained for region D at $t_2$, with
      \Icor\ given in Table~\ref{tab:tot_rads} and the scaled spectral
      distribution of the active region (crosses).  Open dots indicate
      values obtained with the flare spectral distribution described
      Sec.~\ref{sec:models:jcor:spectrum}.
    The thick line indicates the observed value and the dotted line
    its estimated uncertainty. The thin line indicates the continuum level
    for D$_3$.
    \label{fig:lines_abund}
  }
\end{figure}

To explore the effect of the helium chromospheric abundance, we computed a
series of models for kernel~D at $t_2$, using values of \AbHe=0.085, 0.075,
0.065, and 0.055 for $T>6300$~K, i.~e. in the atmospheric layers that strongly
contribute to the formation of the considered lines.  Since a change in the
helium abundance affects the hydrostatic equilibrium, models with the same $T$
vs.~$h$ structure but different abundances give, in principle, different
emitted profiles not only for the He lines, but for all the other
chromospheric lines as well, and for \Halpha\ and the \ion{Ca}{2} lines in
particular.  Therefore, for each value of \AbHe\ we had to compute a different
model.

In Fig.~\ref{fig:lines_abund} we show the computed radiance of the 58.4 He
line and the central intensity of the D$_3$, 1083 and \CaK\ lines,
for the best-matching models for every value of the assumed \AbHe.  The
observed values and their uncertainties, as described in Sec.~\ref{sec:obs},
are also shown.  The variations of \Halpha\ around the mean value of its
central intensity are of the same order as the analogous variations of the Ca
II K line, and thus are not shown in Fig.~\ref{fig:lines_abund}.

The \CaK\ and the \ion{He}{1} 1083 lines depend very weakly on \AbHe, in the
sense that, for each value of \AbHe, it was possible to find a model matching
the observed profiles within the errors.

On the contrary, the \ion{He}{1} D$_3$ and 58.4 lines strongly depend on
\AbHe: the 58.4 radiance is higher than the observed value (beyond the error
bar) for $\AbHe = 0.1$, and quickly decreases for smaller values of \AbHe,
crossing the mean observed value at $\AbHe \sim 0.07$.  On the other hand, the
D$_3$ central intensity increases with decreasing \AbHe, and for \AbHe\
smaller than $\sim 0.075$ significantly exceeds the observed value.  In
practice, the D$_3$ central intensity approaches the photospheric continuum,
i.~e.\ the line \emph{disappears}, for $\AbHe < 0.05$.
  
In other words, both the \ion{He}{1} D$_3$ and 58.4~nm lines become in fact
\emph{weaker} as \AbHe\ decreases.  This result may appear obvious after a
superficial analysis neglecting the non-linearity of non-LTE problems. In
fact, the \ion{He}{1} 1083~nm line does not display the same behaviour: over
the range of abundances explored, it was nearly always possible to find a
temperature structure fitting both that line and the other chromospheric lines
-- with the exception of the lowest value, \AbHe=0.05.

Considering the error of the measurements, we believe that $\AbHe=0.075$ is
the best compromise that allows reasonably good matches for all the considered
lines.  We notice that a variation of $\pm 14$\% of \AbHe\
  (i.~e.: $0.65 \leq \AbHe \leq 0.85$) still allows a good match between
observed and computed lines within the error of the observations.

\begin{figure}[h!tb]
  \centering
  \includegraphics%
  [width=1.0\linewidth]%
  {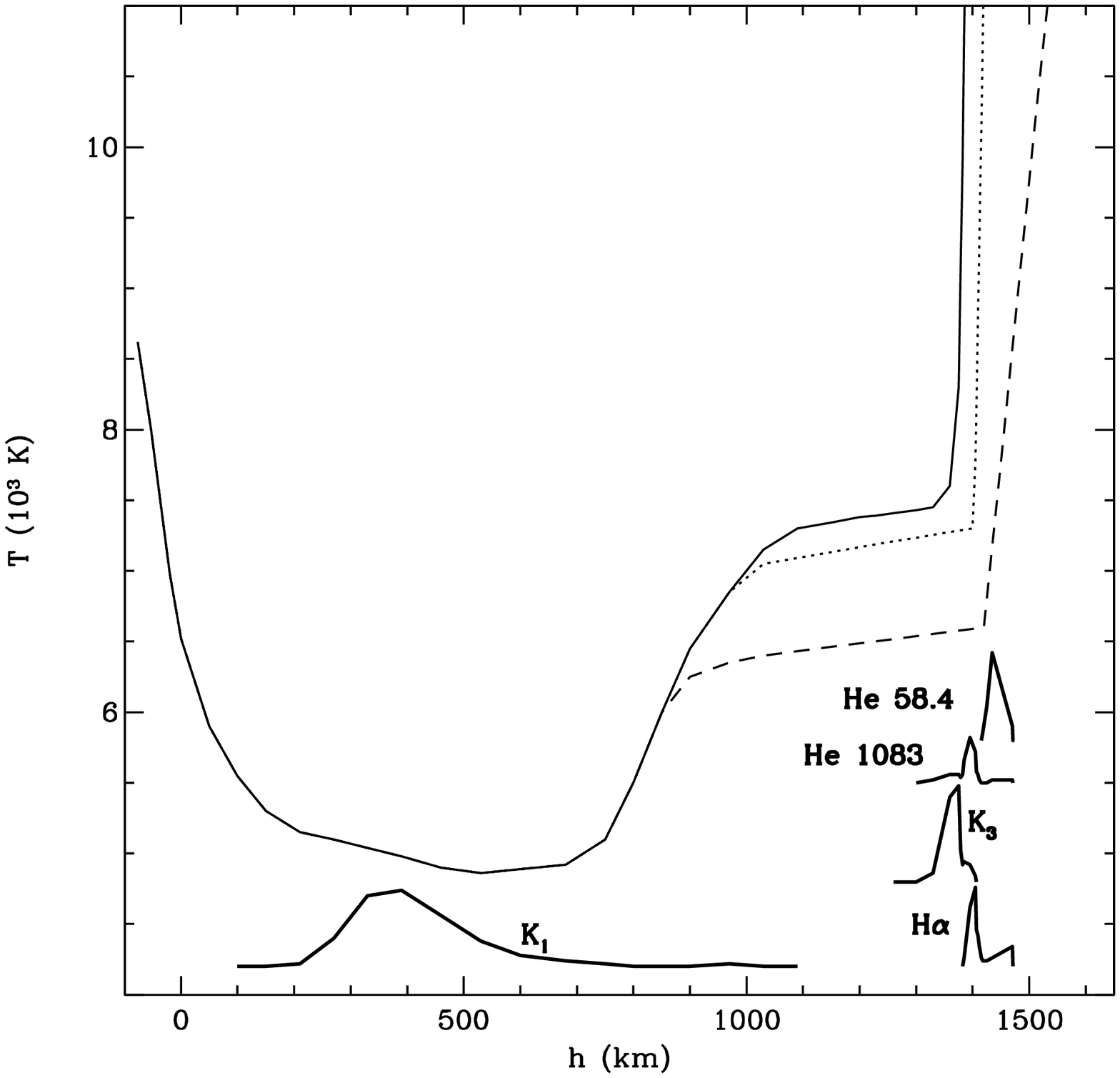}
  \caption{%
    Region D: temperature versus height distribution of the atmospheric models
    obtained for \AbHe=0.075 at the considered times (dotted line: 16:15~UT,
    solid line: 16:20~UT, dashed line: 16:35~UT).  Central
      intensity contribution functions of various lines are also shown, as in
      Fig.~ \ref{fig:mod_kernel_A}.
    \label{fig:mods_kernel_D}
  }
\end{figure}

With this value of \AbHe, we also determined the models for region D at $t_1$
and $t_3$~UT, which are shown in Fig.~\ref{fig:mods_kernel_D}.  The same
figure also shows, for the line cores, the intensity contribution functions
for the model corresponding to the second maximum of EUV emission, at $t_2$.

\begin{figure*}[h!bt]
  \centering
  \includegraphics%
  [angle=90,width=1.0\linewidth]%
  {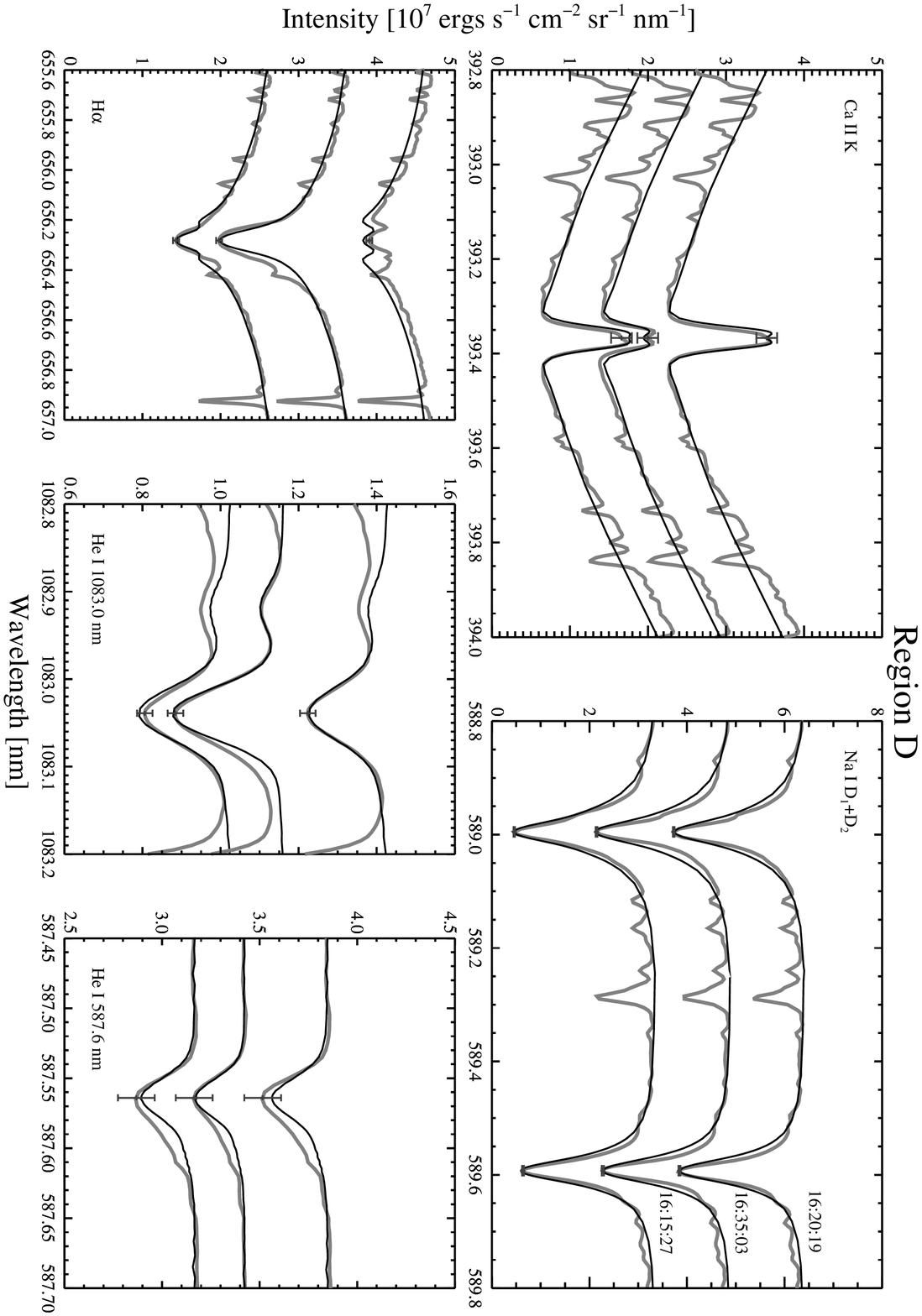}
  \caption{%
    Region~D: comparison between observed (thick grey line) and computed
    profiles (black line).  The profiles at 16:35 and 16:20 UT
    are offset by an arbitrary, constant amount along the ordinate.%
    \label{fig:comp_profiles_D}
  }
\end{figure*}

A comparison between observations and calculations for the three models is
shown in Fig.~\ref{fig:comp_profiles_D}.
The bars in the Figure indicate the uncertainties of the observed profiles.
It can be seen that the agreement found is very good, well within the errors,
with the same caveat regarding the error bars of the optical \ion{He}{1} lines
discussed in Sec.~\ref{sec:models:atmo:A}.
The computed radiances for the UV lines are compared with the
observation in Table~\ref{tab:UV_flux}.  Also in this case, the agreement is
well within the errors in the observations.

\begin{table}[h!tb]
  \begin{center}
    \begin{tabular}{clcccc}
      \tableline
      \tableline
Region & Time      & \multicolumn{2}{c}{\ion{He}{1} 58.4 nm} & \multicolumn{2}{c}{\ion{He}{2} 30.4 nm} \\
       & (UT)  & Obs.      & Comp.     & Obs.      & Comp.    \\        
      \tableline
A      & 16:14 &  2.65     &  2.77     & 16.0      & 15.3     \\
       & 16:33 &  2.65     &  2.91     & 16.0      & 14.0     \\
      \tableline
D      & 16:15 &  4.34     &  4.35     & 32.7      & 31.5     \\
       & 16:20 &  4.93     &  5.29     & 24.0      & 22.9     \\
       & 16:35 &  3.24     &  3.32     & 14.6      & 15.4     \\
      \tableline
    \end{tabular}
  \end{center}
  \caption{%
    Observed and computed radiances for the He UV
    lines. Units of 10$^4$ ergs s$^{-1}$ cm$^{-2}$ sr$^{-1}$.
    \label{tab:UV_flux}
  }
\end{table}

\subsection{The effect of coronal illumination}\label{sec:models:atmo:jcor}

In the discussion on the effect of \AbHe\ on our calculations, we have assumed
the values of \Icor\ given in Table~\ref{tab:tot_rads}.  Those values,
however, have an uncertainty of about 30\%, as estimated in
Sec.~\ref{sec:models:jcor}.  Moreover, the spectral distribution of the
photoionizing photons is also somewhat uncertain.  To assess how these
uncertainties affect the profiles of the helium lines, we performed a set of
calculations for Kernel D at $t_2$, since it corresponds to the second
emission peak (after the initial impulsive peak), when the differences with
the situation studied in Paper~II are the largest.

To test the effect of the uncertainty in the overall coronal
back-illumination, we computed the He profiles increasing and decreasing by
the $\approx 30$\% estimated uncertainty (Sec.~\ref{sec:models:jcor}) the
value listed in Table~\ref{tab:tot_rads}, for the models with
\AbHe=0.1, 0.085, 0.075.
The results of this tests are synthetically shown in
Fig.~\ref{fig:lines_abund} by thin error bars: higher values of \Icor\ produce
stronger \ion{He}{1} lines (deeper optical lines, and higher radiances in the
EUV line).

For \AbHe=0.1, we found that the computed radiances of the \ion{He}{1}~58.4~nm
line change very little.  In fact, a change in \Icor\ much larger than 30\%
(which, we recall is a ``safe'' 3-$\sigma$ value) would be required to match
this radiance within the error bar of the observed value.

For \AbHe=0.075, for which we obtained the best overall fit
(Sec.~\ref{sec:models:atmo:abund})
we found that modifying the value of \Icor\ by 30\% does indeed produce
significant changes in the calculated profiles.
It appears, in this case, that the lower value of \AbHe\ increases the
penetration depth of EUV coronal photons, towards regions where the P-R
mechanism effectively competes with collisional processes in the formation of
the helium spectrum, even of the EUV resonance \ion{He}{1} line.

To assess to what extent the uncertainty in the spectral distribution of EUV
coronal photons can affect the results described in
Sec.~\ref{sec:models:atmo:abund}, we computed the \ion{He}{1} profiles using
the \DEM-derived flare spectral distribution described in
Sec.~\ref{sec:models:jcor:spectrum}, for the cases with \AbHe=0.1, 0.085, 0.075.
The results are indicated in Fig.~\ref{fig:lines_abund} with open dots.

It can be seen that the changes are small, well within the error of the
observations.  We can therefore exclude that the discrepancy between
calculations and observations in the case of \AbHe=0.1, noted in
Sec.~\ref{sec:models:atmo:D}, can be due to the uncertainty in the
photoionization radiation.

In order to explore more systematically the effect of the spectral
distribution of the photoionizing radiation, we also performed tests similar
to the ones included in Paper~II, considering for the spectral distribution
different step functions, which are non-zero only in a limited spectral range,
and zero outside it.
The wavelength ranges we considered for this tests are: 2.5 -- 5.0 nm, 5 -- 10
nm, 10 -- 20 nm, 20 -- 30 nm, 30 -- 40 nm and 40 -- 50.4 nm.

\begin{figure}[h!tb]
  \centering
  \includegraphics%
  [trim=0 28 287 28,height=0.8\textheight]%
  {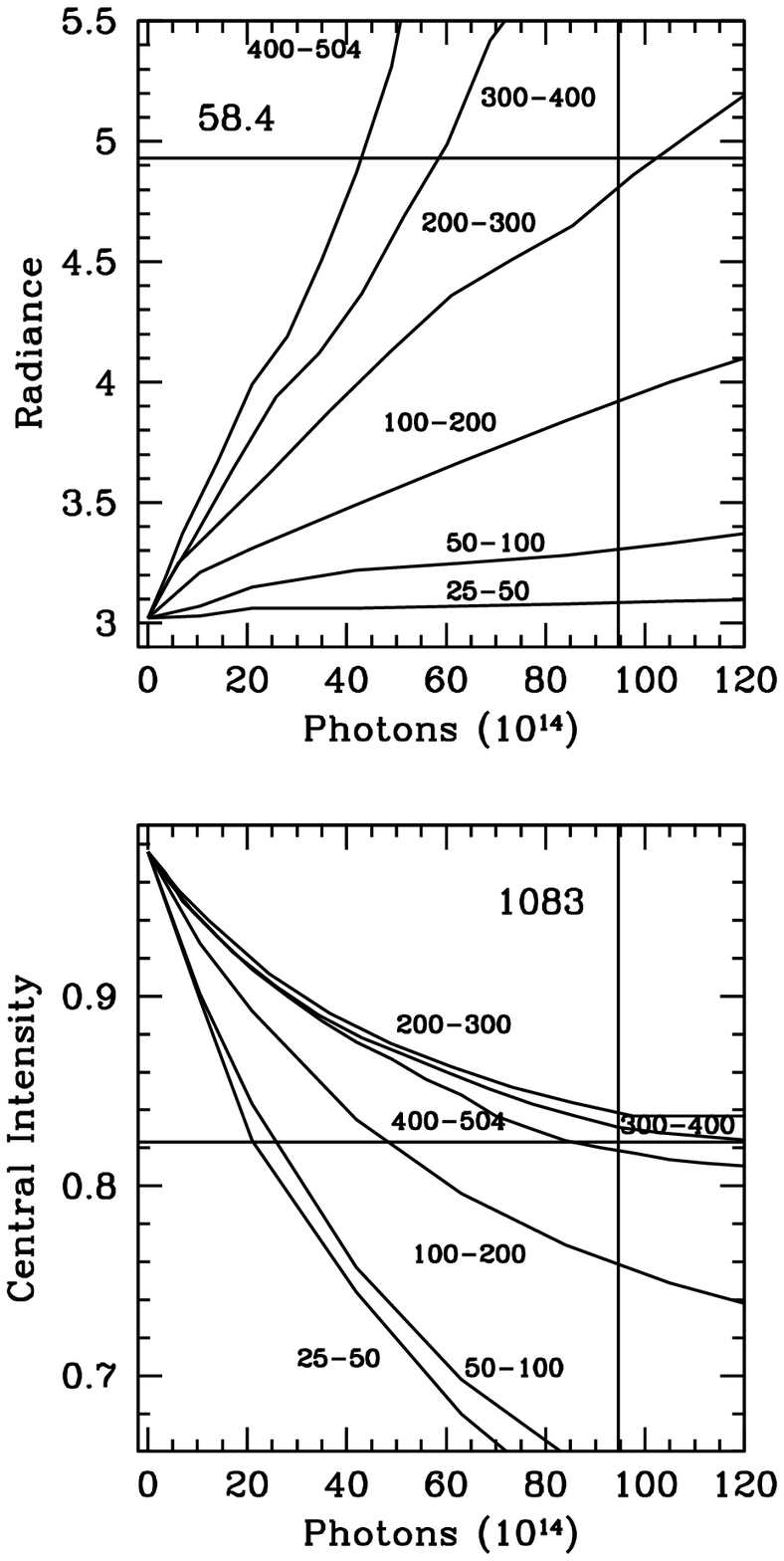}
  \caption{%
    Radiance for the 58.4 nm line (top), and central intensity of the 1083
    line (bottom), as a function of \Icor, for different step-like
    distributions of the field.  Same units as in Fig.~\ref{fig:lines_abund}
    for the ordinates; \Icor\ given in $10^{14}$ ph.\ s$^{-1}$ cm$^{-2}$
    sr$^{-1}$.  The vertical line marks the value of \Icor\ at
      $t_2$, while the horizontal lines indicate the mean value of the
      measured 58.4~nm radiance (top panel) and 1083~nm central intensity
      (bottom panel).
    \label{fig:int_vs_phot}
  }
\end{figure}

In Fig.~\ref{fig:int_vs_phot} we show the intensity in the center of the 1083
line and the radiance of the 58.4 line, as a function of
\Icor. 
We do not show the central intensity of the D$_3$ line,
since its behavior is very similar to the one of the 1083 nm line.
We note that in Paper~II we limited our tests to EUV radiances $\Icor<3\times
10^{15}$ photons s$^{-1}$ cm$^{-2}$ sr$^{-1}$, whereas here we extended the
tests to $1.2\times 10^{16}$.


Several important conclusions can be drawn from this figure. First, the effect
is different on the two lines:
the central intensity of the 1083 line is most sensitive the harder extreme of
the spectrum (down to the threshold of 2.5~nm), while the reverse is true for
the radiance of the 58.4 nm line, which is dominated by the longer wavelength
bands and it is also nearly insensitive to photons below 10~nm.

For the 1083 and the D$_3$ lines the relevant parameter, as in the case
discussed in Paper~II, is the ratio between the helium and hydrogen
photo-ionization cross sections, which in the band 0.5--10~nm is 3 to 5 times
larger than at 50 nm. Therefore, coronal photons at larger wavelengths are
absorbed before they reach the chromospheric heights were these lines are
formed. Since, as we explained in Paper~II, the ratio between the \ion{He}{1}
and H cross sections changes slowly at long wavelengths, the details of the
spectral distribution between 20 and 50.4 nm are of little relevance for the
resulting 1083 nm central intensity.

While these results are qualitatively consistent with those of Paper~II, the
effect of the shortest wavelength photons (in the band 2.5--5.0~nm) is
remarkably different.  While that band had practically no effect on the
\ion{He}{1}~1083 line computed in the AR model discussed in Paper~II, we found
that, in this flaring atmosphere, these photons may have a role in the
formation of that line.

The reason for the difference lies in the higher temperature and densities of
the model chromosphere of region D: the \ion{He}{1}~1083 forms in a region
where most neutral or singly-ionized metals, whose inner-shell absorption
cross sections would otherwise dominate the opacity in that band, have already
mostly disappeared.

The \ion{He}{1}~58.4 nm line, on the other hand, is formed higher in the
transition region, where most hydrogen is ionized, and therefore in this case
the relevant parameter is the helium photo-ionization cross section alone,
which is larger close to the continuum head, at 50.4 nm.

Note that, by contrast with the situation studied in Paper~II, in this case
the much larger coronal radiation has some effect on the radiance of the
\ion{He}{1} UV line, for \AbHe\ up to 0.1 (Fig.~\ref{fig:lines_abund}). The
\ion{He}{2} UV line, on the other hand, which forms at higher temperatures, is
not affected by the coronal radiation, as was the case for the active region
prior to the flare (see Paper~II).

\section{Conclusions}\label{sec:conclusions}

During a small two-ribbon flare, we obtained simultaneous and cospatial
observations, including the chromospheric lines Ca~{\sc ii} K, H$\alpha$ and
Na~{\sc i} D as well as the \ion{He}{1} lines at 587.6 (D$_3$), 1083 and 58.4
nm\ and the \ion{He}{2} line at 30.4 nm.  The EUV irradiance in the ranges
$\lambda<50$~nm and $26<\lambda<34$~nm has also been measured at the same
time.

We analysed the formation of the He lines after the first impulsive burst of
the flare, i.~e. after 16:13 UT, in two different locations of the flaring
area, named A and D. In region A the chromospheric radiance is almost constant
during the time interval we considered, while in region D it approximately
follows the EUV light curve with a maximum emission at about 16:20 UT
(Fig.~\ref{fig:hxr}).  A similar behaviour is also detected for the TR and
coronal lines with temperatures up to $10^6$~K.  We determined the incident
photoionizing EUV radiation produced by the corona overlying the target
regions, by using information from various SOHO instruments (imagers and
spectrographs) and we concluded that in region A the variability of coronal
radiance during the flare is less than a factor 2, while in region D is about
a factor 5 (Table~\ref{tab:tot_rads}).

We determined the spectral distribution of the coronal radiation over the
region~D at 16:20~UT, where we observe a peak of total EUV emission, 
as well as a peak in the hot
\ion{Fe}{19} line (Fig.~\ref{fig:lc_reg}).
We found that the largest differences between the pre-flare and the flare
spectrum occur in the region below 15~nm.

These differences, however, are not large enough to compensate for the
overwhelming contribution of longer wavelength photons to the total EUV
photoionizing emission in this relatively small flare.  We therefore simply
scaled the spectral distribution of a ``typical'' active region by the
appropriate factors listed in the last column of Table~\ref{tab:tot_rads}.
With this coronal radiation, we built semiempirical atmospheric models trying
to match our set of observables for regions A and D, using the same value of
the helium abundance (\AbHe=0.1) we used in Paper~II.  The He lines and
continua are self-consistently computed in the radiative transfer
calculations.

For region A we were successful in obtaining line spectral
intensities and radiances in good agreement, within the
  errors, with the observed ones at the considered times, using a
constant atmosphere and the coronal radiations 
determined in Sec.~\ref{sec:models:jcor}.  Taking into account
  that the formation of the helium lines in this case still falls in a regime
  where the P-R mechanism is important (Fig.~\ref{fig:comp_profiles_A}), this
  result makes us confident that our assumptions for the coronal radiation
and its spectral distribution are acceptable.

For region D, we were able to find models producing profiles reasonably
matching the observations at 16:15 and 16:35, but it was not possible to
obtain a model capable of simultaneously reproduce all the profiles observed
at 16:20, when the flare emission is maximum.  
In particular, it was not possible to simultaneously match \Halpha\ and the
\ion{He}{1}~58.4 line.  We therefore explored the two aspects that most
likely can have an effect on our calculations:
the chromospheric helium abundance,
\AbHe\ (Sec.~\ref{sec:models:atmo:abund}), and the spectral distribution and
overall level of EUV coronal back-illumination
(Sec.~\ref{sec:models:atmo:jcor}).

Since a change in the helium abundance affects the hydrostatic equilibrium,
for each considered value of \AbHe\ we had to compute a different model. We
used values of \AbHe=0.085, 0.075, 0.065 and 0.055 for $T>6300$~K and we
showed that the \ion{He}{1} D$_3$ and 58.4 lines strongly depend on \AbHe,
both becoming weaker as \AbHe\ decreases. In particular the D$_3$ line
disappears for $\AbHe < 0.05$, setting a lower limit for the \AbHe\ value. On
the contrary, over the range of abundances explored, it was always possible to
find an atmospheric structure matching \ion{He}{1} 1083 and the other
chromospheric lines.  Taking into account the error of the measured lines, we
regard $\AbHe=0.075$ as the best compromise that allows reasonably good
matches for all the considered lines and we found that a variation of
$\pm 0.010$ (or: $\pm 14$\%) of \AbHe\ is still acceptable
within the error of the observations, \emph{with the average value of coronal
  illumination determined in Sec.~\ref{sec:models:atmo:jcor}}.

On the other hand, an analysis of the effect of the coronal illumination
\emph{at a given value of \AbHe} (in Sec.~\ref{sec:models:atmo:jcor} we
considered \AbHe=0.1, 0.085 and 0.075) led us to exclude that uncertainties in
the determination of this parameter could be responsible for the discrepancies
at \AbHe=0.1.

We also determined new models for region D at 16:15 and 16:35~UT using
\AbHe=0.075.  Thus, for these two times we have models matching all the
observations for both $\AbHe=0.075$ and $\AbHe=0.1$.  In these cases, as in
the case of Region~A (Sec.~\ref{sec:models:atmo:A}) the incomplete overlap
between the formation regions of \Halpha\ and \ion{He}{1}~58.4 (see for
instance Fig.~\ref{fig:mod_kernel_A}) does not allow a determination of \AbHe\
with an uncertainty better than a factor of two, as in Paper~II.

In summary, our analysis of the observed chromospheric spectra, 
is consistent with \AbHe=0.075; furthermore, we were able to estimate the
uncertainty range as 0.065--0.085 in at least one favorable
case.  The value \AbHe=0.085 -- i.~e.\ the ``canonical'' photospheric helium
abundance \citep[e.~g.:][]{Asplund-etal:05}) -- could
therefore be marginally consistent with our data,
since it results in He features which agree with the observations at the limit
of our estimated errors.

A final note on measuring the helium abundance in the solar chromosphere: The
set of lines we have used, together with an estimate of the EUV coronal
back-radiation, is the \emph{minimum necessary} to estimate \AbHe\ in
chromosphere.  No meaningful estimates, for instance, can be obtained using
only the optical \ion{He}{1} lines or the EUV resonance lines alone.  But (as
in Paper~II, or in all cases considered here, except region~D at 16:20~UT),
even this set of lines may not be \emph{sufficient} for an accurate
determination of \AbHe.  In particular, while the optical \ion{He}{1} lines
are usually well constrained by typical chromospheric lines such as \CaK\ or
\Halpha, the resonance \ion{He}{1} 58.4 line must also be constrained by lines
forming in the lower transition region.  In the high pressure environment of
region~D at the secondary peak of the EUV flare emission, this role is played
by \Halpha.  {In other circumstances, additional constraints provided
  by other lines would be needed: intensities from other upper-chromospheric
  or lower-TR lines (e.~g.: \ion{H}{1} \Lalpha); intensities from other
  \ion{He}{1} and \ion{He}{2} lines (e.~g.: \ion{He}{2} 25.6~nm); spectrally
  resolved profiles of the \ion{He}{1} 58.4 and \ion{He}{2} 30.4 lines. With
  the advent of more recent space missions, such as Hinode and its EUV
  spectrometer (EIS), and with the continued availability of the SOHO
  spectrometers (CDS and SUMER), there is now in fact a window of opportunity
  for obtaining stronger constraints for the kind of modeling described in
  this work. }

\acknowledgements

This work was in part supported by a joint project CNR ({Consiglio
  Nazionale delle Ricerche,} Italy) and CONICET ({Consejo Nacional de
  Investigaciones Cient\`\i ficas y T\'ecnicas,} Argentina).  V.~Andretta also
acknowledges support by the Italian Space Agency (ASI) -- {contracts
  ASI/INAF I/035/05/0 and I/015/07/0}.

We would like to thank the SOHO/CDS and NSO teams for their support in
carrying out the observations analyzed in this paper.  We are also grateful to
R. Falciani and G. Cauzzi who acquired the data at NSO.

SOHO is a project of international cooperation between ESA and NASA.  The NSO
{(National Solar Observatory, USA)} is operated by the Association of
Universities for Research in Astronomy, Inc.\ (AURA), under cooperative
agreement with the National Science Foundation.

We finally thankfully acknowledge the efforts of the CHIANTI consortium
members for making readily available and usable a fine code and database for
astrophysical applications.  
CHIANTI is a collaborative project involving the NRL ({Naval Research
  Laboratory,} USA), RAL ({Rutherford Appleton Laboratory,} UK), MSSL
({Mullard Space Science Laboratory,} UK), the Universities of
Florence (Italy) and Cambridge (UK), and George Mason University (USA).
%




\begin{thebibliography}{31}
\expandafter\ifx\csname natexlab\endcsname\relax\def\natexlab#1{#1}\fi

\bibitem[{{Andretta} {et~al.}(2003){Andretta}, {Del Zanna}, \&
  {Jordan}}]{Andretta-etal:03}
{Andretta}, V., {Del Zanna}, G., \& {Jordan}, S.~D. 2003, \aap, 400, 737

\bibitem[{{Andretta} \& {Jones}(1997)}]{Andretta-Jones:97}
{Andretta}, V., \& {Jones}, H.~P. 1997, \apj, 489, 375

\bibitem[{{Asplund} {et~al.}(2005){Asplund}, {Grevesse}, \&
  {Sauval}}]{Asplund-etal:05}
{Asplund}, M., {Grevesse}, N., \& {Sauval}, A.~J. 2005, in Astronomical Society
  of the Pacific Conference Series, Vol. 336, Cosmic Abundances as Records of
  Stellar Evolution and Nucleosynthesis, ed. T.~G. {Barnes}, III \& F.~N.
  {Bash}, 25--38

\bibitem[{{Avrett} \& {Loeser}(1984)}]{Avrett-Loeser:84}
{Avrett}, E.~H., \& {Loeser}, R. 1984, in {Methods in Radiative Transfer}, ed.
  W.~{Kalkofen} (Cambridge: Cambridge Univ. Press), 341--380

\bibitem[{{Cook} {et~al.}(1994){Cook}, {Keenan}, {Harra}, \&
  {Tayal}}]{Cook-etal:94}
{Cook}, J.~W., {Keenan}, F.~P., {Harra}, L.~K., \& {Tayal}, S.~S. 1994, \apj,
  429, 924

\bibitem[{{Del Zanna} {et~al.}(2002){Del Zanna}, {Landini}, \&
  {Mason}}]{DelZanna-etal:02}
{Del Zanna}, G., {Landini}, M., \& {Mason}, H.~E. 2002, \aap, 385, 968

\bibitem[{{Delaboudini{\`e}re} {et~al.}(1995){Delaboudini{\`e}re}, {Artzner},
  {Brunaud}, {Gabriel}, {Hochedez}, {Millier}, {Song}, {Au}, {Dere}, {Howard},
  {Kreplin}, {Michels}, {Moses}, {Defise}, {Jamar}, {Rochus}, {Chauvineau},
  {Marioge}, {Catura}, {Lemen}, {Shing}, {Stern}, {Gurman}, {Neupert},
  {Maucherat}, {Clette}, {Cugnon}, \& {van Dessel}}]{Delaboudiniere-etal:95}
{Delaboudini{\`e}re}, J.-P. {et~al.} 1995, \solphys, 162, 291

\bibitem[{{Dere} \& {Cook}(1979)}]{Dere-Cook:79}
{Dere}, K.~P., \& {Cook}, J.~W. 1979, \apj, 229, 772

\bibitem[{{Drake}(2003)}]{Drake:03}
{Drake}, J.~J. 2003, Adv. Space Res., 32, 945

\bibitem[{{Falchi} \& {Mauas}(1998)}]{Falchi-Mauas:98}
{Falchi}, A., \& {Mauas}, P.~J.~D. 1998, \aap, 336, 281

\bibitem[{{Feldman} {et~al.}(2005){Feldman}, {Landi}, \&
  {Laming}}]{Feldman-etal:05}
{Feldman}, U., {Landi}, E., \& {Laming}, J.~M. 2005, \apj, 619, 1142

\bibitem[{{Fludra} \& {Schmelz}(1999)}]{Fludra-Schmelz:99}
{Fludra}, A., \& {Schmelz}, J.~T. 1999, \aap, 348, 286

\bibitem[{{Fredvik} \& {Maltby}(1999)}]{Fredvik-Maltby:99}
{Fredvik}, T., \& {Maltby}, P. 1999, \solphys, 184, 113

\bibitem[{{Geiss}(1982)}]{Geiss:82}
{Geiss}, J. 1982, \ssr, 33, 201

\bibitem[{{Geiss}(1998)}]{Geiss:98}
------. 1998, \ssr, 85, 241

\bibitem[{{Gun{\'a}r} {et~al.}(2007){Gun{\'a}r}, {Heinzel}, {Schmieder},
  {Schwartz}, \& {Anzer}}]{Gunar-etal:07}
{Gun{\'a}r}, S., {Heinzel}, P., {Schmieder}, B., {Schwartz}, P., \& {Anzer}, U.
  2007, \aap, 472, 929

\bibitem[{{Harrison} {et~al.}(1995){Harrison}, {Sawyer}, {Carter}, {Cruise},
  {Cutler}, {Fludra}, {Hayes}, {Kent}, {Lang}, {Parker}, {Payne}, {Pike},
  {Peskett}, {Richards}, {Culhane}, {Norman}, {Breeveld}, {Breeveld}, {Janabi},
  {McCalden}, {Parkinson}, {Self}, {Thomas}, {Poland}, {Thomas}, {Thompson},
  {Kjeldseth-Moe}, {Brekke}, {Karud}, {Maltby}, {Aschenbach}, {Brauninger},
  {Kuhne}, {Hollandt}, {Siegmund}, {Huber}, {Gabriel}, {Mason}, \&
  {Bromage}}]{Harrison-etal:95}
{Harrison}, R.~A. {et~al.} 1995, \solphys, 162, 233

\bibitem[{{Hovestadt} {et~al.}(1995){Hovestadt}, {Hilchenbach}, {Burgi},
  {Klecker}, {Laeverenz}, {Scholer}, {Grunwaldt}, {Axford}, {Livi}, {Marsch},
  {Wilken}, {Winterhoff}, {Ipavich}, {Bedini}, {Coplan}, {Galvin}, {Gloeckler},
  {Bochsler}, {Balsiger}, {Fischer}, {Geiss}, {Kallenbach}, {Wurz}, {Reiche},
  {Gliem}, {Judge}, {Ogawa}, {Hsieh}, {Mobius}, {Lee}, {Managadze}, {Verigin},
  \& {Neugebauer}}]{Hovestadt-etal:95}
{Hovestadt}, D. {et~al.} 1995, \solphys, 162, 441

\bibitem[{{Killie} {et~al.}(2005){Killie}, {Lie-Svendsen}, \&
  {Leer}}]{Killie-etal:05}
{Killie}, M.~A., {Lie-Svendsen}, {\O}., \& {Leer}, E. 2005, \apjl, 632, L155

\bibitem[{{Landi} {et~al.}(2006){Landi}, {Del Zanna}, {Young}, {Dere}, {Mason},
  \& {Landini}}]{Landi-etal:06}
{Landi}, E., {Del Zanna}, G., {Young}, P.~R., {Dere}, K.~P., {Mason}, H.~E., \&
  {Landini}, M. 2006, \apjs, 162, 261

\bibitem[{{Mandzhavidze} {et~al.}(1997){Mandzhavidze}, {Ramaty}, \&
  {Kozlovsky}}]{Mandzhavidze-etal:97}
{Mandzhavidze}, N., {Ramaty}, R., \& {Kozlovsky}, B. 1997, \apjl, 489, L99

\bibitem[{{Mandzhavidze} {et~al.}(1999){Mandzhavidze}, {Ramaty}, \&
  {Kozlovsky}}]{Mandzhavidze-etal:99}
------. 1999, \apj, 518, 918

\bibitem[{{Mauas} {et~al.}(2005){Mauas}, {Andretta}, {Falchi}, {et~al.}}]{Mauas-etal:05}
{Mauas}, P.~J.~D., {Andretta}, V., {Falchi}, A., {Falciani}, R., {Teriaca}, L.,
  \& {Cauzzi}, G. 2005, \apj, 619, 604 (Paper II), arXiv:astro-ph/0412058

\bibitem[{{Mauas} {et~al.}(1997){Mauas}, {Falchi}, {Pasquini}, \&
  {Pallavicini}}]{Mauas-etal:97}
{Mauas}, P.~J.~D., {Falchi}, A., {Pasquini}, L., \& {Pallavicini}, R. 1997,
  \aap, 326, 249

\bibitem[{{Murphy}(2007)}]{Murphy:07}
{Murphy}, R.~J. 2007, \ssr, 130, 127

\bibitem[{{Pietarila} \& {Judge}(2004)}]{Pietarila-Judge:04}
{Pietarila}, A., \& {Judge}, P.~G. 2004, \apj, 606, 1239

\bibitem[{{Share} \& {Murphy}(1998)}]{Share-Murphy:98}
{Share}, G.~H., \& {Murphy}, R.~J. 1998, \apj, 508, 876

\bibitem[{{Teriaca} {et~al.}(2003){Teriaca}, {Falchi}, {et~al.}}]{Teriaca-etal:03}
{Teriaca}, L., {Falchi}, A., {Cauzzi}, G., {Falciani}, R., {Smaldone}, L.~A.,
  \& {Andretta}, V. 2003, \apj, 588, 596 (Paper I)

\bibitem[{{Vernazza} \& {Reeves}(1978)}]{Vernazza-Reeves:78}
{Vernazza}, J.~E., \& {Reeves}, E.~M. 1978, \apjs, 37, 485

\bibitem[{{Young} {et~al.}(2003){Young}, {Del Zanna}, {Landi}, {Dere}, {Mason},
  \& {Landini}}]{Young-etal:03}
{Young}, P.~R., {Del Zanna}, G., {Landi}, E., {Dere}, K.~P., {Mason}, H.~E., \&
  {Landini}, M. 2003, \apjs, 144, 135, astro-ph/0209493

\bibitem[{{Zirin}(1975)}]{Zirin:75}
{Zirin}, H. 1975, \apjl, 199, L63

\end{thebibliography}


\end{document}